\PassOptionsToPackage{unicode=true}{hyperref} % options for packages loaded elsewhere
\PassOptionsToPackage{hyphens}{url}
\PassOptionsToPackage{dvipsnames,svgnames*,x11names*}{xcolor}
\documentclass[]{article}
\usepackage{lmodern}
\usepackage{amssymb,amsmath}
\usepackage{ifxetex,ifluatex}
\usepackage{fixltx2e} % provides \textsubscript
\ifnum 0\ifxetex 1\fi\ifluatex 1\fi=0 % if pdftex
  \usepackage[T1]{fontenc}
  \usepackage[utf8]{inputenc}
  \usepackage{textcomp} % provides euro and other symbols
\else % if luatex or xelatex
  \usepackage{unicode-math}
  \defaultfontfeatures{Ligatures=TeX,Scale=MatchLowercase}
\fi
\IfFileExists{upquote.sty}{\usepackage{upquote}}{}
\IfFileExists{microtype.sty}{%
\usepackage[]{microtype}
\UseMicrotypeSet[protrusion]{basicmath} % disable protrusion for tt fonts
}{}
\usepackage{xcolor}
\usepackage{hyperref}
\hypersetup{
            pdftitle={Ranked Sparsity: A Cogent Regularization Framework for Selecting and Estimating Feature Interactions and Polynomials},
            colorlinks=true,
            linkcolor=Maroon,
            filecolor=Maroon,
            citecolor=Blue,
            urlcolor=blue,
            breaklinks=true}
\usepackage[margin=1in]{geometry}
\usepackage{graphicx,grffile}
\makeatletter
\def\maxwidth{\ifdim\Gin@nat@width>\linewidth\linewidth\else\Gin@nat@width\fi}
\def\maxheight{\ifdim\Gin@nat@height>\textheight\textheight\else\Gin@nat@height\fi}
\makeatother
\setkeys{Gin}{width=\maxwidth,height=\maxheight,keepaspectratio}
\ifx\paragraph\undefined\else
\let\oldparagraph\paragraph
\renewcommand{\paragraph}[1]{\oldparagraph{#1}\mbox{}}
\fi
\ifx\subparagraph\undefined\else
\let\oldsubparagraph\subparagraph
\renewcommand{\subparagraph}[1]{\oldsubparagraph{#1}\mbox{}}
\fi

\makeatletter
\def\fps@figure{htbp}
\makeatother

\usepackage{amsfonts}
\usepackage{float}
\usepackage{caption}
\usepackage{gensymb}
\usepackage[nomarkers,figuresonly,nofiglist]{endfloat}
\usepackage[nodisplayskipstretch]{setspace}
\usepackage[margin = 1in]{geometry}
\usepackage{indentfirst}
\usepackage[bottom]{footmisc}
\usepackage{booktabs}
\usepackage{longtable}
\usepackage{array}
\usepackage{multirow}
\usepackage{wrapfig}
\usepackage{float}
\usepackage{colortbl}
\usepackage{pdflscape}
\usepackage{tabu}
\usepackage{threeparttable}
\usepackage{threeparttablex}
\usepackage[normalem]{ulem}
\usepackage{makecell}
\usepackage{xcolor}

\title{Ranked Sparsity: A Cogent Regularization Framework for Selecting and
Estimating Feature Interactions and Polynomials}
\author{}
\date{\vspace{-2.5em}}

\begin{document}
\maketitle

\noindent  Ryan A. Peterson, Joseph E. Cavanaugh \newline

\noindent  \textbf{Abstract} \newline 

\noindent  We explore and illustrate the concept of ranked sparsity, a
phenomenon that often occurs naturally in modeling applications when an
expected disparity exists in the quality of information between
different feature sets. Its presence can cause traditional and modern
model selection methods to fail because such procedures commonly presume
that each potential parameter is equally worthy of entering into the
final model -- we call this presumption ``covariate equipoise''.
However, this presumption does not always hold, especially in the
presence of derived variables. For instance, when all possible
interactions are considered as candidate predictors, the premise of
covariate equipoise will often produce over-specified and opaque models.
The sheer number of additional candidate variables grossly inflates the
number of false discoveries in the interactions, resulting in
unnecessarily complex and difficult-to-interpret models with many (truly
spurious) interactions. We suggest a modeling strategy that requires a
stronger level of evidence in order to allow certain variables
(e.g.~interactions) to be selected in the final model. This ranked
sparsity paradigm can be implemented with the sparsity-ranked lasso
(SRL). We compare the performance of SRL relative to competing methods
in a series of simulation studies, showing that the SRL is a very
attractive method because it is fast, accurate, and produces more
transparent models (with fewer false interactions). We illustrate its
utility in an application to predict the survival of lung cancer
patients using a set of gene expression measurements and clinical
covariates, searching in particular for gene-environment
interactions.\newline


\noindent  \textbf{Declarations}: Not applicable. \newline

\noindent  \textbf{Date last modified}: 12/06/2021 \newline

\noindent  \textbf{Availability of data and material}: Data used in this
application is publicly available via GEO database accession number
GSE68465. \newline

\noindent  \textbf{Code availability}: Code for simulations and methods
is included as supplemental material.\newline

\begin{center}\rule{0.5\linewidth}{0.5pt}\end{center}

\noindent  Ryan Peterson
\newline Department of Biostatistics \& Informatics, University of
Colorado School of Public Health, Aurora, CO \newline email:
ryan.a.peterson@cuanschutz.edu (corresponding author) \newline

\noindent  Joseph Cavanaugh
\newline Department of Biostatistics, University of Iowa College of
Public Health, Iowa City, IA \newline email: joe-cavanaugh@uiowa.edu
\newline

\doublespacing
\newpage

\hypertarget{introduction}{%
\section{Introduction}\label{introduction}}

In the ever-growing, ever-changing field of model selection and machine
learning, ``black-box'' predictive models (e.g.~neural networks) have
become increasingly popular (and increasingly opaque). When one's
exclusive desire is predictive accuracy, these difficult-to-interpret
models are often worth a certain lack of understanding. However, overly
complex predictive contexts are not generally compatible with the
traditional aim of science: to explain and to understand the world in
which we live. With their growing popularity, black-box models are
starting to be applied in situations where explanation \emph{should} be
the primary goal. Worse yet, in some circumstances, there is little
regard for the consideration that more transparent models could produce
similar prediction results. So, as scientists continually increase the
number of candidate predictors, those building models are using
increasingly complicated functions of candidate predictors in order to
optimize for predictive performance above all else. Is there a
justifiable way to hold on to the traditional aims of science amid these
trends?

In this paper, we will argue that the benefits reaped from choosing a
black-box model must be weighed against the interpretative costs of a
lack of scientific understanding. However, before we can proffer a
method to accomplish this goal, we must first answer a salient question
-- why do black-box methods outperform transparent models in prediction?
The answer is difficult because these black-box methods are diverse, as
are the situational considerations that make a particular method perform
better or worse. Broadly speaking, the benefits of black-box methods can
be roughly explored by investigating situations where transparent linear
models fail. We will focus in particular on the issue of bias caused by
model misspecification.

Say we have a vector of data made up of a response of interest
\(\boldsymbol y\) and a matrix of covariates (or predictors) \(X\), some
columns of which are related to \(\boldsymbol y\) while others are not.
Suppose that the variates comprising \(\boldsymbol y\) are independent
(conditional on \(X\)) and can be conceptualized as following a
distribution in the exponential family. One can envision many ways of
fitting an optimal predictive model to \(\boldsymbol y\), but a popular
method (if transparency is a goal) is to fit generalized linear models
(GLMs) based on all possible subsets of covariates to select the best
model on the basis of an information criterion. This method is somewhat
limited to lower-dimensional settings, because the number of candidate
models increases combinatorically with the dimension of \(X\). However,
in recent times the Least Absolute Shrinkage and Selection Operator (the
lasso) has changed the landscape surrounding the problem of identifying
a suitable predictive model (Tibshirani, 1996). With the lasso and its
many extensions, it is possible to have an extremely high-dimensional
covariate space and still end up with a relatively well-fit,
interpretable model. In either setting, if the true generating model has
informative interactions among covariates and/or meaningful nonlinear
covariate-response relationships, black-box methods exist that can
outperform even the best traditional or lasso model (given enough data).
This is because none of the candidate main effect (i.e.~transparent)
models can capture the model's complex interaction/polynomial terms; all
of the candidate transparent models are misspecified.

One potential solution, given the power and flexibility of the lasso,
would be simply to add ``derived variables'' of \(X\), such as
interactions and polynomials, into a new (potentially very large) design
matrix. The lasso \emph{can} simultaneously select and estimate
important interactions and polynomials even in this ultra-high
dimensional setting. However, in this paper, we will show that this
method yields a preponderance of both false and missed discoveries,
unless proper methods are used to incorporate what we call ``ranked
sparsity.''

The paper is organized as follows. First, we intuitively motivate the
concept of ranked sparsity, illustrating its necessity when looking for
active interactions. Second, we propose the sparsity-ranked lasso, and
connect it to some other related concepts and regularization methods
that have been proposed in the literature, as well as other
state-of-the-art interaction selection methods. Next, we present
simulation studies to investigate the performance of the SRL compared to
competing methods in the polynomial and interaction selection setting.
We then apply the SRL in a high-dimensional setting of gene-environment
interaction selection in the context of a lung cancer application.
Finally, we discuss the strengths and weaknesses of the SRL relative to
other strategies that have been proposed.

\hypertarget{ranked-sparsity}{%
\section{Ranked Sparsity}\label{ranked-sparsity}}

\hypertarget{intuitive-motivation}{%
\subsection{Intuitive Motivation}\label{intuitive-motivation}}

Ranked sparsity, which we also refer to as ranked skepticism, is a
philosophical framework that challenges the traditional implementation
of Occam's Razor in the context of variable selection. In Einstein's
words\footnote{Debate exists regarding whether this is a true quote or a paraphrase of Einstein.},
the maxim stipulates that ``everything must be made as simple as
possible, but not simpler.'' This is a noble goal, but some obvious
questions arise: how do we know when a model is as simple as it should
be? How should we measure simplicity in the first place? Specifically,
we wish to challenge the ubiquitous answers to these questions in the
field of model selection, which rely on a presumption that we call
``covariate equipoise'': the prior belief that all covariates are
equally likely to enter into a model. To illustrate this idea, say we
are trying to find a well-fit model to predict an outcome
\(\boldsymbol y\) using a set of covariates, including age, weight, and
height. Of the candidate models below, which is ``simpler''?
\begin{align}
E(y_i) &= \beta_0 + \beta_1 \text{Age}_i + \beta_2 \text{Weight}_i + \beta_3 \text{Age}_i*\text{Weight}_i  \label{eq:a}\\
E(y_i) &= \beta_0 + \beta_1 \text{Age}_i + \beta_2 \text{Weight}_i + \beta_3 \text{Height}_i \label{eq:b}
\end{align} Virtually all variable selection tools assume these two
models to be equally simple. Due to the presumption of covariate
equipoise, simplicity is equated to parsimony, and is measured only by
the number of parameters in the model (which is 4 in both models).
However, any statistician would quickly recognize that model
\eqref{eq:b} is an order of magnitude easier to understand and
communicate than model \eqref{eq:a}. We argue that a proposed model's
simplicity should therefore not only be tied to its level of parsimony,
but also to its transparency as measured by the ease at which it can be
understood and communicated (a metric loosely tied to the number of
interactions and nonlinear terms in the model). Ultimately, a good model
interpreted correctly is better than a great model interpreted
erroneously. This concept is the primary motivation for the ranked
sparsity methods we introduce in this paper, which provide a means of
searching for important interactions/nonlinear terms without rendering
the chosen model unnecessarily opaque.

In this work, for the linear form used to characterize the mean outcome,
we use the term ``main effects'' to refer to the regression coefficients
on the original covariates of interest, and ``interaction effects'' to
refer to coefficients on the product of covariates that correspond to
the main effects. We also define the ``sparsity level'' as the
proportion of candidate variables (a.k.a. features, covariates, or
predictors) that are inactive in a given true generating model. A high
sparsity level thus indicates that a smaller proportion of candidate
variables are truly important. Conversely, the ``saturation level'' is
defined as the proportion of candidate variables that are active. The
sparsity level is governed by a mix of what cannot be known about
nature's true generating model and what can (sometimes) be known about
the ambition of a particular scientific project. In sparse settings,
consistent model selection criteria such as the Bayesian Information
Criterion (BIC) (Schwarz, 1978) and its extensions (Bogdan et al., 2008;
Chen and Chen, 2008) have been shown to be effective, while in saturated
settings, efficient criteria such as Cp, AIC, and corrected AIC
(Mallows, 1973; Akaike, 1974; Hurvich and Tsai, 1989) perform relatively
well. This difference in the performance of various model selection
criteria suggests that in settings where multiple levels of sparsity are
to be expected among different groups of covariates, the optimal
criterion needs to account for this disparity in some way by penalizing
the covariates differently. The ``ranking'' that occurs in our concept
of ranked sparsity thus refers to settings where the sparsity levels
within covariate groups are expected to be ordered in a specific way
\emph{a priori}.

With this impetus in mind, given a saturation level in the main effects,
we can show that the maximum saturation level attainable for the
(first-order) interaction effects is limited by ``hierarchy''
assumptions about the true generating model. Sometimes called model
heredity or the marginality principle, model hierarchy refers to the
rules pertaining to which interactions can be nonzero, and it is
typically broken down into ``strong,'' e.g.
\(E(y_i) = \beta_0 + \beta_1 x_{1i}+ \beta_2 x_{2i} + \beta_3 x_{1i}*x_{2i}\);
``weak,'' e.g.
\(E(y_i) = \beta_0 + \beta_1 x_{1i}+ \beta_3 x_{1i}*x_{2i}\); and
``anti-'' (or ``non-'') hierarchical models, e.g.
\(E(y_i) = \beta_0 + \beta_3 x_{1i}*x_{2i}\). As an illustration of how
hierarchy limits saturation, consider a case where only 3 of 30 possible
main effects are active, then strong hierarchy would dictate that in the
generating model, only \(\binom{3}{2} = 3\) signal variables can exist
in the interaction set. Under weak hierarchy, then this quantity is
limited to \(\sum_{j=1}^3 (30 - j) = 84\) active interactions. In either
case, the number of signals in the interaction set is bounded by the
number of signal variables in the main effects, as are their saturation
(sparsity) levels (see supplemental materials for proofs). Of course,
outside of simulation settings, the hierarchy status of the generating
model will be unknown. Even so, there is good reason to believe that the
saturation level in one set of covariates (interaction effects) is going
to be less than the saturation level in another group of covariates
(main effects). As a result, it becomes necessary to account for this
disparity somehow in the model selection process; we cannot simply apply
the same penalty to both main effects and interactions and expect
optimal performance.

\hypertarget{the-sparsity-ranked-lasso}{%
\subsection{The Sparsity-Ranked Lasso}\label{the-sparsity-ranked-lasso}}

In this section, we propose and motivate the Sparsity-Ranked Lasso (SRL)
as a tool for implementing ranked sparsity in the search for important
derived variables of a feature space. Suppose we have \(p\) features
\(\left[x_1, x_2, ..., x_p\right] = X_{n \text x p}\), and a centered
response variable \(\boldsymbol y\); some (but not all) features are
related to \(\boldsymbol y\). For this section, we assume that the
variates comprising \(\boldsymbol y\) are normally distributed and
conditionally independent given \(X\); however, it will become clear
that the concept and development apply more generally in the GLM family.
The lasso (Tibshirani, 1996) has become immensely popular in this
setting for its computational efficiency and its effectiveness in
variable selection. The lasso simultaneously estimates coefficients for
each of the \(p\) features and selects from them, such that they are
either ``active'' (i.e. \(\hat \beta_j \neq 0\)), or ``inactive''
(\(\hat \beta_j = 0\)). The estimated nonzero coefficients do suffer
from a bias that is introduced by the lasso's penalty term, but this
bias is often warranted as it significantly attenuates the variance
associated with having too saturated of a model. Typically, the
magnitude of shrinkage induced by the lasso's penalty term is treated as
a tuning parameter (\(\lambda\)) and selected on the basis of an
information criterion or cross-validation (CV). In this section, we will
show how the lasso is expected to fail when applied to feature sets of
different sizes, most notably when applied to interactions and main
effects, and we will offer a solution via the sparsity-ranked lasso.

The ordinary lasso solution can be obtained by standardizing all of the
columns of \(X\) and minimizing the following expression with respect to
\(\boldsymbol \beta\): \[
\left|\left|\boldsymbol y - X\boldsymbol \beta\right|\right|^2 + \lambda \sum_{j=1}^p | \beta_j|
\] It is well-known that this solution has a Bayesian interpretation. If
each \(\beta_j \sim \text{Laplace}(0, \lambda)\), then the mode of the
joint posterior distribution represents the lasso solution (Tibshirani,
1996) for a given \(\lambda\) value. As the sample size increases, the
likelihood becomes more concentrated and contributes more information to
the posterior, eventually pulling the mode off of zero. As \(\lambda\)
is increased, the balance of information shifts toward the Laplace
prior, and the mode gets pulled (potentially all the way) to zero (a
visualization of this is available at
\url{https://ph-shiny.iowa.uiowa.edu/rpterson/shiny_vis1/}). These
zero-centered independent Laplace priors form the following joint prior
density: \[
\pi (\boldsymbol \beta) = 
\prod_{j=1}^p \frac {\lambda}{2 } e^{-\lambda |\beta_j|}
\]

As a brief aside, we turn to the concept of Fisher information. Fisher
information is invoked in likelihood theory to describe the behavior of
maximum likelihood estimators, but the concept can quantify the
structural characteristics of any joint density.
\footnote{Jeffreys, for instance, derived his famous noninformative prior based on the concept of the Fisher information of a prior density.}
For \(W \sim f(w | \lambda)\) where \(\lambda \in \Lambda\) is scalar
and \(\lambda \rightarrow \log f(w | \lambda)\) is twice differentiable
in \(\lambda\) for every \(w\), the model Fisher information at any
\(\lambda\) is defined to be
\(I(\lambda) = E_{W|\lambda} \left[-\frac {\partial^2}{\partial \lambda^2} \log f(W|\lambda)\right]\).

Now, consider partitioning the covariate space \(X\) into \(K\) groups,
such that \(X = \left[A_1, A_2, ..., A_k, ..., A_K\right]\). If we let
\(p_k\) refer to the column dimension of \(A_k\) \(\forall \ k\), and
let \(\beta_j^k\) refer to a particular \(\beta_j\) in covariate group
\(k\), then the prior for \(\boldsymbol \beta\) undergoes a purely
cosmetic change and becomes \[
\pi (\boldsymbol \beta | \lambda) \propto
\prod_{k=1}^K \prod_{j = 1}^{p_k} \lambda  e^{-\lambda |\beta_j^k|}
\]

\noindent If we think of all of the \(\beta_j^k\) as random variables
(which they are \emph{a priori}) and take \(\lambda\) to be a parameter,
it becomes straightforward to find the Fisher information in this prior
density. \[
\begin{aligned}
\frac {\partial^2}{\partial \lambda^2}  \log \pi (\boldsymbol \beta | \lambda) =
- \frac{1}{\lambda^2} \sum_{k=1}^K p_k
\\
I(\lambda) = E_{X|\lambda} \left[
-\frac {\partial^2}{\partial \lambda^2} \log \pi (\boldsymbol \beta | \lambda)
\right] 
 = \frac{1}{\lambda^2} \sum_{k=1}^K p_k
\end{aligned}
\]

\noindent This information increases with the dimension of each group's
parameter space equally (for any \(\lambda > 0\)). So if \(p_1 = 10\),
and \(p_2 = 100\), by default the contribution toward the prior
information by the covariate group \(A_1\) is only one tenth that of the
covariate group \(A_2\), for any \(\lambda > 0\). If \(A_1\) refers to
the main effects, and \(A_2\) refers to their pairwise interactions,
there can be a substantial degree of \emph{a priori} informational
asymmetry between the interactions and the main effects. This asymmetry
leads popular feature selection tools such as the ``all-pairwise lasso''
(APL) to select too many selections among the candidate interaction
effects while shrinking the main effects excessively.

In many (perhaps most) situations, the preceding weighting scheme may
not be desired. We can slightly modify the prior distribution by
replacing \(\lambda\) with \(\lambda_k = \lambda \sqrt{p_k}\). Now,
unlike before when the distributions were independent and identical for
all \(k\), each \(\beta_j^k\) is only independent and identically
distributed within its own covariate group \(k\). The Fisher information
contained in the prior for covariates in group \(k\) after this
modification is \[
I(\lambda_k) = \frac{p_k}{\lambda^2_k} = \frac{p_k}{\lambda^2 p_k} = \frac{1}{\lambda^2} \ \forall \ k
\]

\noindent  In words, by scaling each group's penalty by the square-root
of its dimension, we have ensured that the prior information is the same
across groups; no group has an \emph{a priori} informational advantage.
Therefore, we can achieve a ``ranking'' in the sparsity that treats
covariate \emph{groups} equally as opposed to the covariates themselves.
If we add another tuning parameter, \(\gamma\), in the definition for
\(\lambda_k\) such that \(\lambda_k = \lambda p_k^\gamma\), the
resulting approach can be seen as a generalization the ordinary lasso,
where \[
I(\lambda_k) = \frac {1}{\lambda^2} p_k^{(1-2\gamma)}
\]

\noindent  If \(\gamma = 0\), this is identical to the ordinary lasso.
If \(\gamma = 0.5\), then \(w_k =\sqrt {p_k}\) and each covariate group
contributes the same amount of prior information (which is a good
default setting for many circumstances, especially for the context of
interactions). As \(\gamma\) increases, the penalties for larger groups
of covariates increase quickly (as the information contribution
decreases quickly with group size). In less-common cases where the
grouping is not well-defined, we suggest tuning \(\gamma\) to a value
between zero (the ordinary lasso) and 0.5 (equal group-level prior
information). Under this primary SRL formulation, we do not suggest
choosing \(\gamma > 0.5\) unless there is a strong reason to believe
that the quality of information decreases substantially with group size.

We consider a variant of the penalty weighting scheme where the
information is expected to decrease as the \emph{group index} increases.
Specifically, instead of weighting the penalties by \(p_k^{1-2\gamma}\),
we use \(({\sum_{i=1}^k p_i})^{1-2\gamma}\). This group index
formulation yields a group-level information contribution of \[
I(\lambda_k) = \frac {1}{\lambda^2}\left(\frac{p_k}{\sum_{i=1}^k p_i}\right)^{(1-2\gamma)}
\]

\noindent  In words, the information contribution is highest for
\(A_1\), and decreases cumulatively as more groups are added. If \(A_1\)
represents main effects, and \(A_2\) represents squared polynomial terms
of those main effects, this cumulative group index penalty ensures that
the polynomial terms are penalized more heavily than the main effects
(despite having the same group size). We call this formulation the
cumulative SRL, for which the value of \(\gamma\) determines the extent
to which penalty increases with the group index (e.g.~polynomial order).
We suggest using cross-validation to tune \(\gamma\) for the cumulative
SRL.

With these changes, the objective function resembles closely that for
the adaptive lasso (Zou, 2006), minimizing the following with respect to
\(\boldsymbol \beta\): \[
\left|\left|\boldsymbol y - X\boldsymbol \beta\right|\right|^2 + \lambda \sum_{k=1}^{K} \sum_{j=1}^{p_k}  w_k| \beta_j^k|
\]

\noindent Unlike the adaptive lasso, \(w_k\) only depends on the group
dimensions and potentially \(\gamma\) and the group index. Specifically,
\(w_k = p_k^{(1-2\gamma)}\) for the original SRL (\(w_k = \sqrt{p_k}\)
assuming group parity in prior information), and
\(w_k = \left(\frac{p_k}{\sum_{i=1}^k p_i}\right)^{(1-2\gamma)}\) for
the cumulative SRL. As has been shown in other work (e.g. Wang and Wang,
2014), this objective function can be slightly modified to handle
non-normal outcomes (binary, Poisson, survival, etc.) by substituting
the negative log-likelihood for the least squares term, minimizing
\(-l(\boldsymbol \beta) + \lambda \sum_{k=1}^{K} \sum_{j=1}^{p_k} w_k| \beta_j^k|\)
with respect to \(\boldsymbol \beta\); such a substitution was employed
for this paper's application. We suggest using path-wise coordinate
descent to optimize this objective function with respect to
\(\boldsymbol \beta\); this can be implemented with either the
\texttt{sparseR} package or \texttt{ncvreg} (Breheny and Huang, 2011).

To summarize, while the ordinary lasso presumes throughout its path that
the sparsity levels are equal among covariate groups, the SRL can
enforce a ranking in the expected sparsity levels such that the amount
of contributed prior information is controlled across covariate groups.
We also introduced a ``cumulative'' variant of the SRL, which is
particularly useful for selecting and estimating polynomial effects. We
illustrated how either the primary SRL or its cumulative variant can be
tuned with \(\gamma\). A sensible choice for the SRL is to utilize
\(\gamma=0.5\), which sets the prior information to be equal across all
covariate groups; we suggest fixing \(\gamma=0.5\) when utilizing the
SRL specifically for interactions or for otherwise well-defined
covariate groups (or both, as is the case in this paper's application).
The cumulative SRL performs best when \(\gamma\) is tuned among several
possible values (e.g. \(\gamma \in \{0, .5, 1, 2, 4, 8, ...\}\),
suggesting an increased amount of penalization for higher orders of
polynomials), which can be done using an information criterion or CV. We
explore the performance of the SRL in the forthcoming simulation studies
and application, after a brief discussion of similar ideas and
techniques in the literature.

\hypertarget{related-methods}{%
\subsection{Related Methods}\label{related-methods}}

\hypertarget{existing-methods-which-penalize-based-on-group}{%
\subsubsection{Existing Methods which Penalize Based on
Group}\label{existing-methods-which-penalize-based-on-group}}

Several methods exist that are close in spirit to a general ranked
sparsity framework, including the Integrative Lasso with Penalty Factors
(IPF-lasso) (Boulesteix et al., 2017) and the priority lasso (Klau et
al., 2018). In both methods, each group of covariates has its own
estimated penalty. The IPF-lasso creates a new tuning parameter for each
covariate group \(k\), estimating \(\lambda_k\) using a grid search and
cross-validation. While technically possible for any number of groups,
the IPF-lasso can become computationally difficult when multiple groups
are considered (though notably, an adaptive extension of the IPF-lasso
which mitigates this issue has been proposed and implemented in the
\texttt{ipflasso} R package). Similarly, the priority lasso incorporates
a priority ordering of feature groups and fits sequential lasso models
on these groups, using residuals from each model as a new outcome to
predict using next most important feature set. The priority lasso is
feasible for multiple groups; however in order to avoid over-optimism,
Klau et al. (2018) recommend a cross-validated offset schema which can
be computationally intensive and difficult to implement for multiple
groups.

Another related lasso extension meriting discussion is the sparse group
lasso (SGL). The solution to the SGL is found by minimizing the
following with respect to \(\boldsymbol \beta\): \[
\left|\left|\boldsymbol y - X\boldsymbol \beta\right|\right|^2 + 
\alpha \lambda \sum_{k=1}^K \sum_{j=1}^{p_k} |\beta_j^k| + 
(1-\alpha) \lambda \sum_{k=1}^K \sqrt{p_k} ||\boldsymbol \beta^k ||
\]

\noindent SGL bears a resemblance to the SRL, noticing in particular the
factor multiple of the group-level penalty, \(\sqrt{p_k}\). However,
despite the similar penalty scaling, the SGL will yield quite different
results to the SRL in practice. While SGL shrinks the magnitude of the
entire vector \(\boldsymbol \beta^k\) within each group, SRL penalizes
each coefficient in some sense independently from the others in its
group. For example, if the magnitude of the first coefficient in group
one is large, i.e. \(\hat \beta_1^1 >> 0\), the SRL would not induce any
effect on the magnitude of \(\hat \beta_2^1\). The SGL still penalizes
each variable separately in its first penalty, but its second penalty is
on the group-level magnitude. In our hypothetical example, a large
\(\hat \beta_1^1\) coefficient would thus relax the penalty on
\(\hat \beta_2^1\) to an extent. See Figure 1 in Friedman, Hastie and
Tibshirani (2010) for a good visualization of this principle. A more
detailed comparison of the performance of the SGL to the SRL is left for
future work, but one benefit of the SRL that is already evident is its
lack of the need for an additional tuning parameter, as \(\gamma = 0.5\)
assumes equal prior information across groups, and is thus a defensible
choice for many circumstances.

\hypertarget{existing-methods-for-interaction-selection}{%
\subsubsection{Existing Methods for Interaction
Selection}\label{existing-methods-for-interaction-selection}}

Several methods have been proposed for selecting and estimating
interactions under the weak and/or strong hierarchy ``constraint.'' The
hierNet approach (Bien, Taylor and Tibshirani, 2013) is well-suited for
low-dimensional problems due to its computational complexity. A similar
regularization-based method, glinternet (Lim and Hastie, 2015), has been
shown to be as effective as hierNet in selecting interactions, but able
to execute the fitting and selection 10-10000 times faster. The ``strong
heredity interaction model'' (SHIM) approach is similar to the hierNet
approach; it extends the lasso to select interaction terms while under a
strong hierarchy constraint. SHIM also adds an adaptive lasso element to
achieve the oracle property (Choi, Li and Zhu, 2010), and uses an
IPF-lasso-type approach of tuning the penalty for the interactions
separately from the main effects. SHIM thus has an additional tuning
parameter to cross-validate over. Yet another approach is called
``regularization under marginality principle'' (RAMP) (Hao, Feng and
Zhang, 2018), which is a two-stage regularization approach that is
useful for settings where the storage of the interaction model matrix is
an issue. By having a first-stage screening via regularization on the
main effects, RAMP substantially cuts down on the size of the model
matrix in its second stage, only considering candidate interactions that
made it past the first selection stage. All of these methods constrain
the solution path to weakly or strongly hierarchical models. Another set
of 12 alternative methods for determining treatment-biomarker
interaction screening via various types of regularization and dimension
reduction are described and empirically evaluated in Ternès et al.
(2017).

It is important to note that many of these methodological works on
interaction selection involve a comparison to the APL, which
consistently selects too many interactions in these comparisons. This
issue gets compounded when the true generating model has very few
``active'' interactions, which is an ongoing limitation of interaction
feature selection for some of these methods (Lim and Hastie, 2015). The
SRL, we will show, does not suffer this limitation. Further, one
consideration that is often mentioned only briefly in these related
works is whether or not we \emph{should} restrict all candidate models
to be hierarchical. The usual presumption is that it makes the most
sense for all candidate models to be hierarchical. However, Chipman
(1996) provides a compelling paradigm for model hierarchy in a Bayesian
context, and argues why the strict imposition of hierarchical structures
may not always be defensible. If we think of interactions as children of
their ``parent'' main effects, we would guess that a child is certainly
\emph{most likely} to be in a model if its parents are both in the
model. It is comparably less probable that a child is in a model if one
of its parents is not. Is it absolutely impossible (with probability
zero) for a child to be in a model without either of its parents?

There are numerous occasions where a generating model is not, in fact,
hierarchical. Chipman gives the example of the atmospheric sciences,
where relations of the form \(Y = A\exp(BC)\) are common, which is a
non-hierarchical model on the log scale. We can also point to models for
lung cancer, where ``pack-years,'' the interaction between the number of
years spent smoking and the reported number of cigarette packs smoked
per day, is an acknowledged risk factor on its own. In fact, a
non-hierarchical model of this type is plausible in any setting where
level of exposure and time of exposure are both captured somewhere in
the candidate covariate space. Therefore, we argue that in lieu of
hierarchy constraints, a better general rule would be to enforce
hierarchy \emph{preference}. This is considerably different than a
constraint. We have shown that the SRL enforces higher penalties for
interactions than for the main effects (when \(p > 2\)), which naturally
enables hierarchy preference (but does not force hierarchy).

Finally, it is feasible that the IPF-lasso could be used to select from
all possible interactions by defining main effects and interaction
effects as two separate blocks, an approach which would enforce neither
a hierarchy preference nor a constraint. Such an approach, while
defensible, may suffer from imprecision (in ensuring that interactions
and main effects contribute proportionally to the prior information).
Boulesteix et al. (2017) advise to investigate penalty factors within
each group as \((1, 2^\gamma)\) for a sequence of positive and negative
integers \(\gamma\) (in the two-group case). For interactions, if
candidate \(2^\gamma\) values are not near \(\sqrt{\binom{p}{2}}\), we
would expect asymmetry in the prior information. More generally, similar
imprecision is likely to occur when covariate groups vary substantially
in size. For example, in one application, Boulesteix et al. (2017)
investigate combining clinical data (11 features) with microarray gene
expression measurements (22,283 features) to predict the survival of
patients with breast cancer. Their CV procedure estimated the optimal
penalty factor for the genetic expression data to be \(2^5=32\), thereby
penalizing the expressions much more than the clinical data (in fact,
there were no selected gene expressions). With the SRL, prior to CV, we
know that for the gene-expression measurements to contribute the same
amount of prior information as the clinical features, this penalty
factor should be \(\sqrt{22,283 / 11} \approx 45\). Therefore, despite
being optimized via CV, 32 is likely too low for the genetic features; a
penalty factor of 45 would still select no expression features, in that
sense the model would be the same. However, with a penalty factor of
only 32, the coefficients on the clinical features may be shrunk more
than necessary, thereby decreasing the power to detect clinical effects
as well as (perhaps) the predictive accuracy of the final model.

\hypertarget{the-sparser-package}{%
\subsection{\texorpdfstring{The \texttt{sparseR}
Package}{The sparseR Package}}\label{the-sparser-package}}

We have developed an R package, \texttt{sparseR}, which works in concert
with \texttt{ncvreg} (Breheny and Huang, 2011) to implement and
facilitate the ranked sparsity methods discussed in this paper. By
building upon the \texttt{recipes} package (Kuhn and Wickham, 2019),
\texttt{sparseR} also provides a useful means of preprocessing data sets
before model fitting, which can facilitate the use of a mix of factors,
binary variables, and continuous variables as covariates. The package
also contains an information-criterion based metric that we call RBIC
(Peterson, 2019), which is paired with a forward step-wise selection
function that can select from all possible interactions and polynomials
under strong, weak, or non-hierarchy using the ranked-sparsity
framework. The \texttt{sparseR} package and a detailed tutorial will
soon be made available on the Comprehensive R Archive Network (CRAN). A
development version is available on GitHub at
\url{https://github.com/petersonR/sparseR}.

\hypertarget{simulations}{%
\section{Simulations}\label{simulations}}

\hypertarget{polynomial-simulation-study}{%
\subsection{Polynomial Simulation
Study}\label{polynomial-simulation-study}}

\begin{figure}

{\centering \includegraphics[height=6in]{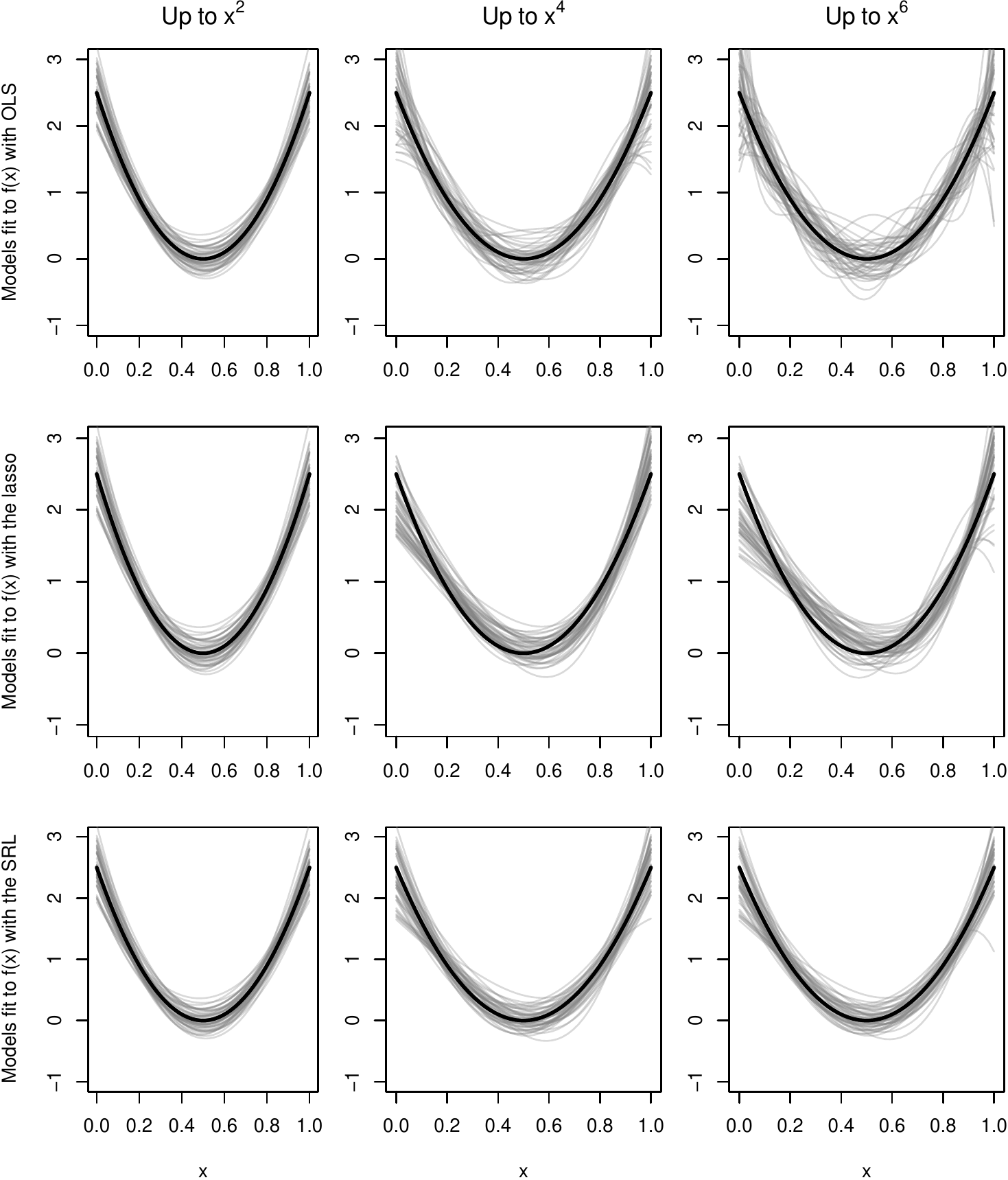} 

}

\caption{A simple simulation where 50 samples of size 100 are generated for $x$ and a response variable $y$ with the relationship $y = f(x) + N(0, .9^2)$.  The black line represents the true $f$, and the grey lines represent 50 fits to the different samples. Models in the top three plots are fit using ordinary least squares (OLS); in the middle three plots, models are fit with the lasso; and on the bottom three plots, models are fit using the sparsity-ranked lasso (SRL). SRL and lasso models are tuned using BIC. The covariates included are the polynomials of $x$ up to $x^2$ (left) up to $x^4$ (center), and up to $x^6$ (right).}\label{fig:Fig_01}
\end{figure}

Consider a simple simulated example, where we have 100 observations
arising from a true \(f(x) = 10*(x-.5)^2\) measured with some residual
noise, \(\varepsilon_i \overset {iid} \sim N(0, 0.9^2)\), and
\(x_i \overset {iid} \sim \text {unif}(0,1)\). This relationship is
shown by the solid black line in Figure
\ref{fig:Fig_01}\footnote{The R language and environment version 4.0.2 is used for all figures in this work.}.
It is well-known that the addition of extraneous polynomial terms in a
regression model hurts the model's predictive performance, especially at
the bounds of the covariate space. This can be seen in the top three
plots of Figure \ref{fig:Fig_01} -- adding higher order terms increases
the ``wiggliness'' of the fits (represented by the grey lines).

With only one covariate, we could easily fit, say, \(m\) models with
increasing orders of polynomials up to \(m\) and select the best order
using an information criterion. However, in higher dimensional settings
with \(p\) covariates, this approach is not practical since the number
of possible models is \(2^{pm}\). In such settings, one might think to
use the lasso to select the optimal order -- this method is explored in
the middle 3 plots of Figure \ref{fig:Fig_01}. Evidently, the bias
incurred by the L1 penalty reduces some of the variability (the
``wiggliness'') in the relationship, while at the same time
contaminating the shape of relationship (note that the fitted lines are
bent down towards the origin).

The cumulative SRL can be successfully applied for such polynomial
models; if we set \(w_j = (d_j)^\gamma\), where \(d_j\) refers to the
degree of covariate \(j\), this approach is equivalent to the cumulative
group-index SRL in the one-covariate case. The resulting fits, where
\(\gamma \in \{0.5, 1, 2\}\) is selected by
BIC\footnote{We exclude $\gamma=0$ in this application of the cumulative SRL to showcase how the method differs from the ordinary lasso applied to the polynomials, which is equivalent to setting $\gamma=0$.},
are shown by the grey lines in the bottom three plots of Figure
\ref{fig:Fig_01}. We observe that the fits are both reducing the
wiggliness from the extraneous terms while at the same time inducing
less ``bending'' towards the origin.

\begin{figure}

{\centering \includegraphics[height=4in]{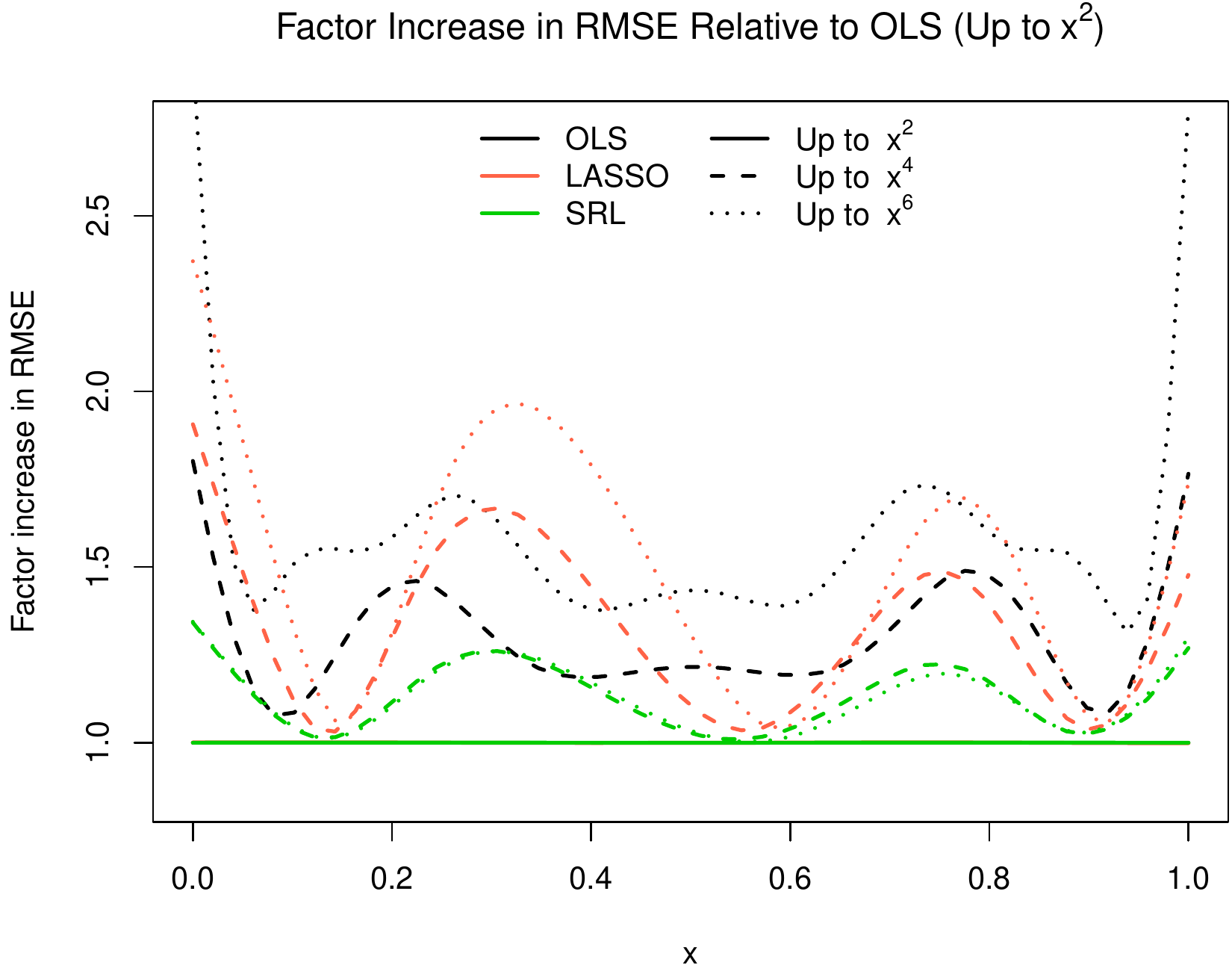} 

}

\caption{The expected increase in the root-mean-squared error (RMSE) for the ordinary least squares (OLS), lasso, and sparsity-ranked lasso (SRL) models relative to a baseline ``oracle'' OLS model that only includes the ``true'' variables ($x$ and $x^2$).}\label{fig:Fig_02}
\end{figure}

The plots in Figure \ref{fig:Fig_01} only show 50 fits each, but
repeating this process 10,000 times, we can compare the
root-mean-squared error (RMSE) of estimation resulting from each method
across the domain of
\(x\)\footnote{The RMSE of estimation is based on the sum of the squared deviations between the true mean values and the corresponding estimates under the fitted model and is computed for 50 evenly-spaced points along the domain of \textit{x}.}.
In Figure \ref{fig:Fig_02}, we show the increase in the RMSE for each
method relative to a baseline ``oracle'' model (i.e.~an ordinary
least-squares model that only includes the ``true'' \(x\) and \(x^2\)).
We find that while there is no replacement for an ``oracle'' model, the
next best models are those which utilize the SRL method. Interestingly,
there appears to be very little in terms of predictive difference
between the SRL applied up to the 4th order and SRL applied up to the
6th order; this implies that we could likely increase the maximum order
and still not observe a substantive impact on the predictive
performance. On the other hand, if the lasso is used, there is a marked
decrease in predictive performance between the 4th order model and the
6th order model; it appears as though these models (as well as the OLS
models) perform increasingly poorly as the number of extraneous
polynomials increases.

We conducted a separate simulation study to further establish the
performance of the cumulative SRL for more general functional forms.
Between kernel and spline methods, many options exist for estimating a
smooth relationship in low-dimensional settings. We investigate four
possible generating models: polynomials of orders 10, 2 (quadratic), 1
(linear), and 0 (null). In each setting, we generate data (\(n=100\))
according to the following model: \[
\begin{aligned}
&y_i = \alpha + \beta_1 x_i + \beta_2 x_i^2 + ... + \beta_{10}x_i^{10} + \varepsilon_i \\
&\text{where } \varepsilon_i \overset{iid}{\sim} N(0,1) \text{ and } x_i \overset{iid}{\sim} \text{unif}(0,1)
\end{aligned}
\]

In order to compare many possible functional forms (for the high-order
and 2-order settings), the \(\beta_j\) parameters were randomly
generated. For the high-order setting, we drew
\(\theta_1, \theta_2, ..., \theta_{10} \sim N(0,1)\), then scaled them
so their magnitude sums to 10;
\(\beta_{j} = 10*\theta_j / \sum |\theta_i|\). The same technique was
used in the quadratic generating model, except only for
\(j \in \{1,2\}\); all other parameters were set to 0. In the linear
case, \(\beta_1=10\), and in the null setting,
\(\beta_j = 0 \ \forall j\). The sole covariate \(x\) follows a standard
uniform distribution within each simulation.

\begin{figure}

{\centering \includegraphics[height=6in]{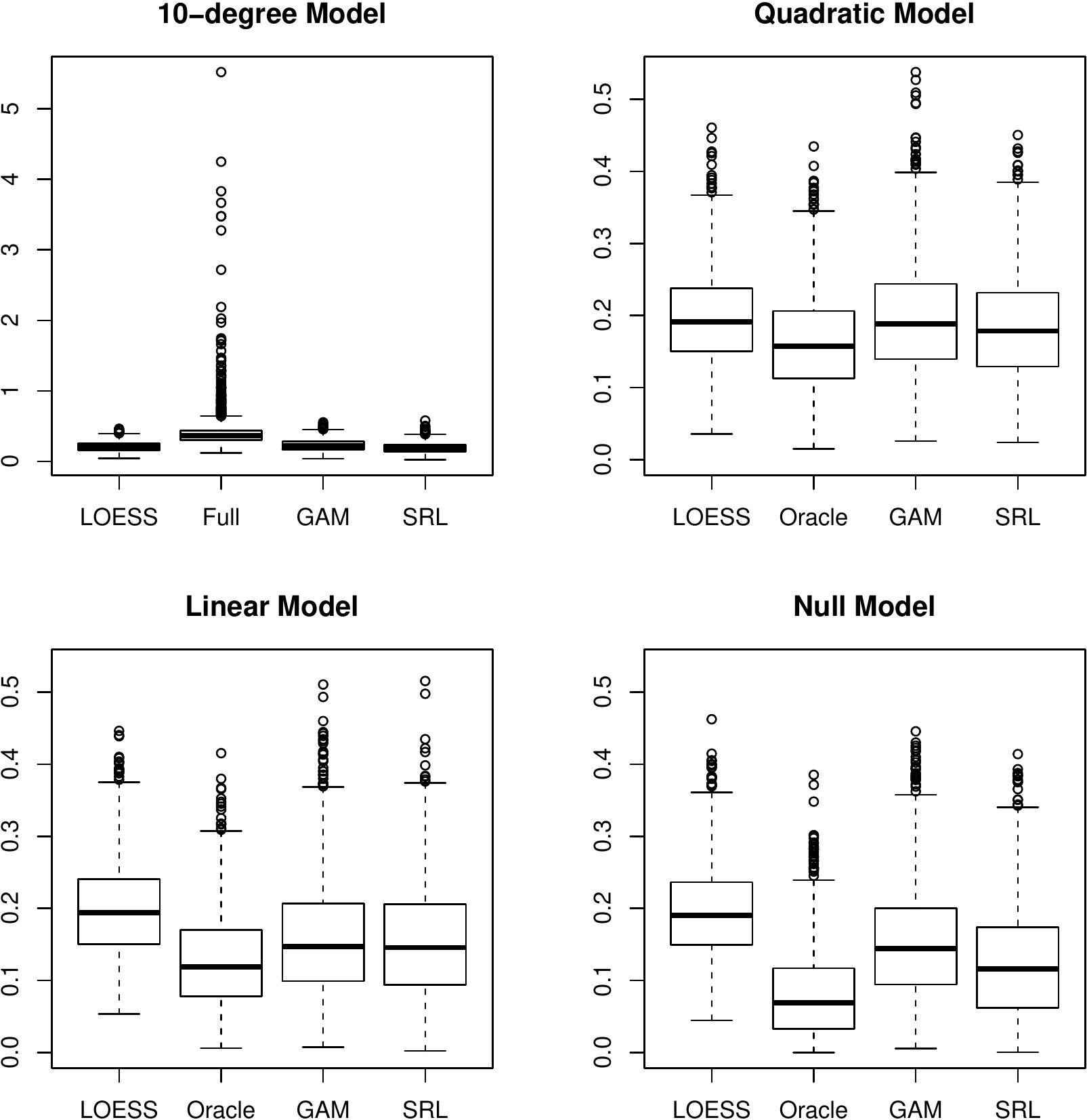} 

}

\caption{Performance of smoothing methods in describing a truly polynomial (or null) relationship between a single covariate and response. The RMSE within each simulation is plotted along the y-axis. SRL and lasso models are tuned using BIC.}\label{fig:Fig_03}
\end{figure}

For model fitting, we utilize the cumulative SRL method with all terms
up to the 10th order. We compare this model fit with the LOESS smoother
(\texttt{loess()} in the \texttt{stats} package (R Core Team, 2020)) and
with a smoothing spline (the \texttt{gam()} and \texttt{s()} functions
from the \texttt{mgcv} package (Wood, 2011)). The default settings are
used for these functions. The cumulative SRL is tuned using repeated
(\(r=5\)) 10-fold cross-validation with
\(\gamma \in \{0, 0.5, 1\}\)\footnote{The $\gamma$ options of 0, 0.5, and 1 represent a minimal set of possible ranked sparsity settings for additional penalization for higher-order polynomials from none ($\gamma=0$) to strong ($\gamma=1$).}.
The simulations are repeated 1,500 times. Models are evaluated on the
basis of the RMSE of estimation on \(n=10,000\) new randomly sampled
observations, presented in Figure \ref{fig:Fig_03}. We find that the
cumulative SRL achives similar performance to its LOESS and spline
alternatives. Importantly, in contrast to alternatives, the cumulative
SRL performs very well for the lower order and null models, predicting
new values almost as well as the oracle model. This relative improvement
is the result of overfitting of the alternative methods. SRL's
superlative performance in the null setting matters to a great extent in
the high-dimensional setting where we expect many null relationships.
For the \(10^{th}\)-order generating model, the poor performance of the
``full'' model, despite it being technically correctly specified, is a
mark of the fact that there is high correlation among the polynomials of
\(x\), which inflates the variance of the estimated regression
coefficients.

These simulation studies suggest that, at least for functional forms
well-represented by polynomials, the cumulative SRL will do well to fit
the relationship while also sifting through many null relationships. For
other functional forms not well-represented by polynomials, or when the
covariate has a high amount of skew, the cumulative SRL will not perform
as well as spline/kernel alternatives. However, for the case of
covariate skew, a normalizing transformation on the covariate prior to
expanding the polynomial can mitigate this effect. We have developed
software in a separate work that can adequately and robustly perform
these normalizations (Peterson and Cavanaugh, 2020; Peterson, 2021).

\hypertarget{interactions-simulation-study}{%
\subsection{Interactions Simulation
Study}\label{interactions-simulation-study}}

\hypertarget{simulation-setup}{%
\subsubsection{Simulation Setup}\label{simulation-setup}}

While we have shown how the SRL can compete with other smoothing
techniques in the one-dimensional setting, the main benefits of the SRL
methodology are present in the medium-to-high dimensional setting where
model selection must take place. We set up a more extensive simulation
in the context of interactions, comparing the SRL's performance to that
of the glinternet method, the all-pairwise lasso (APL), and the lasso
with only the main effects included (LS0).

Let \(\mathbb X = \left[X, X^{\odot 2}\right]\) refer to the column
combination of the main covariates (an \(n \times p\) matrix), followed
by their element-wise product values (\(n \times \binom{p}{2}\)). We
wish to fit the linear model
\(\boldsymbol y = \mathbb X \boldsymbol \beta + \boldsymbol \varepsilon\),
where we partition
\(\boldsymbol {\beta}^T = \left[\boldsymbol {\beta}_1^T, \boldsymbol {\beta}_2^T\right]\)
to correspond with our notation for \(\mathbb X\). In the usual case,
where interactions are not considered, it is assumed that
\(\boldsymbol {\beta}_1\) is the only parameter vector with nonzero
components. One would expect that this assumption helps in situations
where the true generating model is, in fact, linear in the main
covariates with no active interactions. However, what if there are
nonzero components in the other parameter vector? We will investigate.

In the simulations to follow, we take \(n=300\), and \(p=20\), and we
generate each element in \(X\) as independent uniform(0,1) random
variables. The reason for independence in these predictors is to allow
simple interpretations of the selection results (with correlated
predictors, what constitutes a false discovery or a false negative is
less well-defined). We investigate predictive performance in the setting
of correlated features in the supplemental work, and mention the
take-aways in this paper's discussion. We set the number of nonzero main
effects (\(s\)) in \(\boldsymbol \beta_1\) as \(s = 5\). In order to
generate our \(\boldsymbol \beta\) coefficients in such a way that a
large set of possible relationships are considered, we use scaled normal
random variables as our ``active'' (i.e.~nonzero) parameters in
\(\boldsymbol \beta\). Specifically, we consider 11 generative settings
of interest corresponding to the number of active interactions \(b\),
and the algorithm to generate the nonzero (active) coefficients is
comprised of the following steps:

\singlespacing

For \(b \in \{0,1,2,...,10\}\):

\begin{itemize}
\item
  Draw \(\theta_1, \theta_2, ..., \theta_s \sim N(0,1)\)
\item
  Compute scaled main effects
  \(\beta_{1j} = 10*\theta_j / \sum |\theta_i|\)
\item
  Generate \(b\) standard normal variables \(\{\phi_1, ..., \phi_b\}\)
\item
  Select the index of active interactions \(j\) according the rules
  outlined in the following paragraph
\item
  Set \(\beta_{2j} = 10*\sqrt {\frac{12}{7}}\phi_j/\sum |\phi_i|\)
  \footnote{Scaling by $\sqrt {\frac{12}{7}}$ accounts for the difference in variability between uniform random features and their interactions.}
\end{itemize}

\doublespacing

The generating models were not necessarily strongly hierarchical. In
particular, each simulation was configured according to Chipman's
paradigm; strong hierarchy was most probable, weak hierarchy less so,
and non-hierarchy least so. Given \(s = 5\) and \(p=20\), there are
\(\binom{s}{2}=10\) candidate interactions that would yield a strongly
hierarchical model, \(s (p - s)=75\) that would yield a weakly
hierarchical model, and \(\binom{p-s}{2} = 105\) that would yield an
non-hierarchical model. Within a simulation, each active interaction
effect (if there were any) was drawn at random from these bins of
strong, weak, or non-hierarchical candidate effects with probabilities
0.7, 0.2, 0.1, respectively.

In order to fit these models, we consider four modeling frameworks: LS0
(lasso on original terms only), APL (lasso with original and all
pairwise interaction terms), SRL (sparsity-ranked lasso with original
and all pairwise interaction terms), and GLN (glinternet model). For
each framework, the optimal \(\lambda\) is selected with 10-fold CV, and
then that tuned model is used to predict 10,000 new randomly sampled
observations. The \(\gamma\) parameter for SRL is fixed to 0.5,
corresponding to an equal contribution of prior information from the
main effects and the interaction effects. This process (including the
new generation of \(\beta_{ij}\) terms) is repeated 1,000 times in order
to check the models' predictive and selective performance. For the
former, we use the RMSE of prediction on newly generated data to compare
the predictive accuracy of the final
models\footnote{The RMSE of prediction is based on the sum of the squared deviations between each new observation and the corresponding predicted value under the fitted model.}.
For the latter, we use the false discovery rate (FDR), the mean number
of Type I errors, and the mean number of Type II errors, examining these
quantities both collectively and separately for interactions and main
effects. In particular, these selection metrics for interaction terms
loosely measure the ``transparency'' and ``interpretability'' of the
various models; models with many false discoveries in the interaction
effects are needlessly opaque.

\begin{figure}

{\centering \includegraphics[height=3.5in]{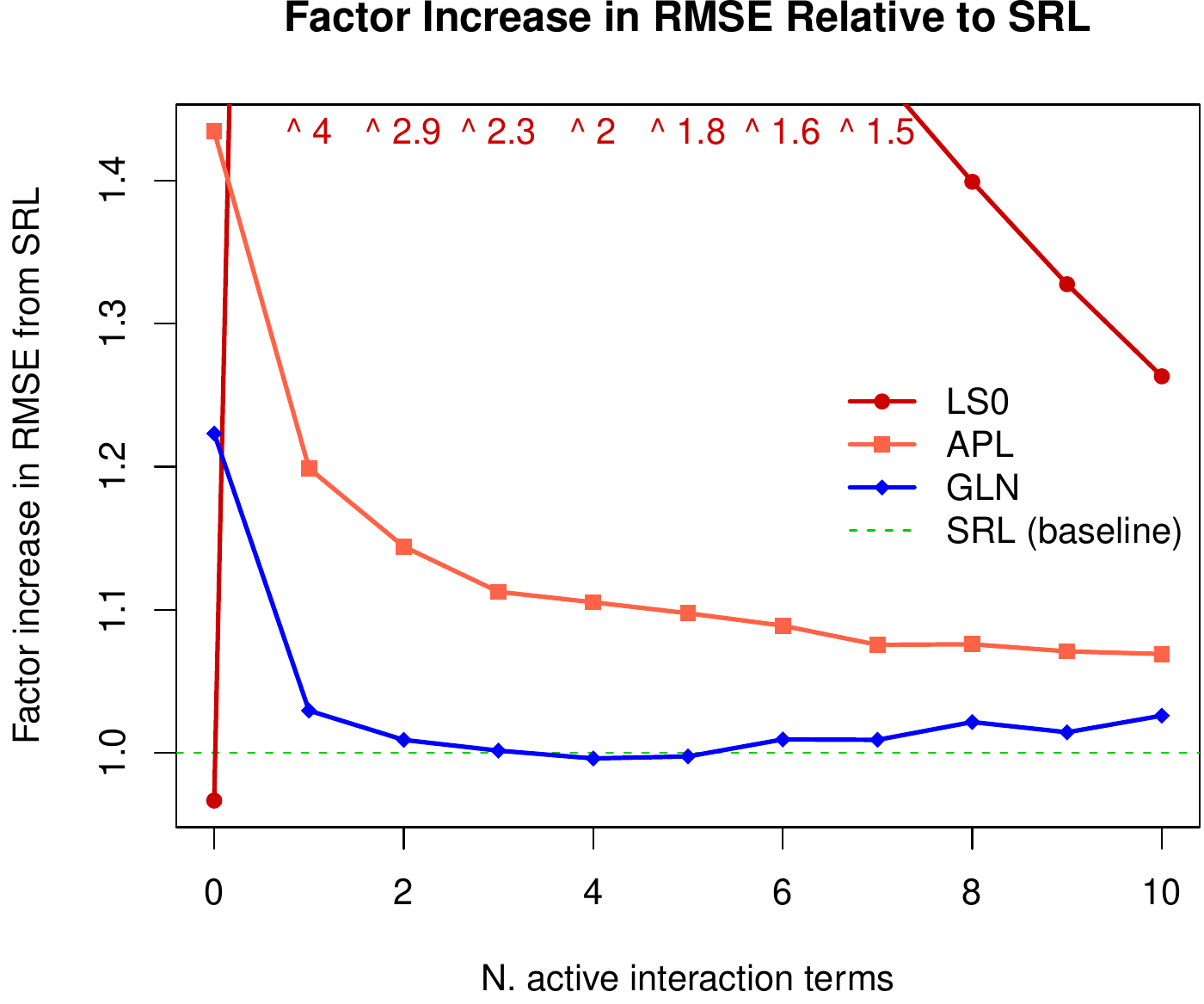} 

}

\caption{Predictive performance of various interaction fitting methods relative to SRL. LS0 refers to the lasso fit using only the original terms, APL refers to the lasso fit using the original terms and all pairwise interactions, SRL refers to the sparsity-ranked lasso fit with $\gamma = 0.5$, and GLN refers to the glinternet model. For all models, $\lambda$ was tuned with 10-fold cross-validation. The ``\textasciicircum'' notation refers to the values of the LS0 model that were too large to be clearly plotted next to the other curves.}\label{fig:Fig_04}
\end{figure}

\hypertarget{simulation-results}{%
\subsubsection{Simulation Results}\label{simulation-results}}

The predictive performance of the models across all simulations is shown
in Figure \ref{fig:Fig_04}. Although the LS0 model demonstrates a very
slight gain in predictive performance if the true model has no
interactions, it also exhibits a marked loss in performance when any
active interactions are present. The APL model performs comparably
better than the LS0 model when any interactions are present, but
performs much worse in the no-interaction case. SRL performs much better
than either the APL or GLN in the no-interaction case. SRL and GLN have
similar predictive performance to each other when active interactions
are present, both performing much better than either the APL or LS0. In
the supplement, we show how other correlation structures exhibit similar
results; in fact, SRL's relative performance compared to GLN and APL is
sometimes even better with higher correlation among features. In
situations where the SRL performed worse than other methods (such as the
compound symmetry correlation matrix with feature correlation
\(\rho = 0.5\)), selecting the optimal \(\gamma \in \{0, 0.5, \infty\}\)
using CV still performed comparably to the best alternative.

\begin{figure}

{\centering \includegraphics[height=4.2in]{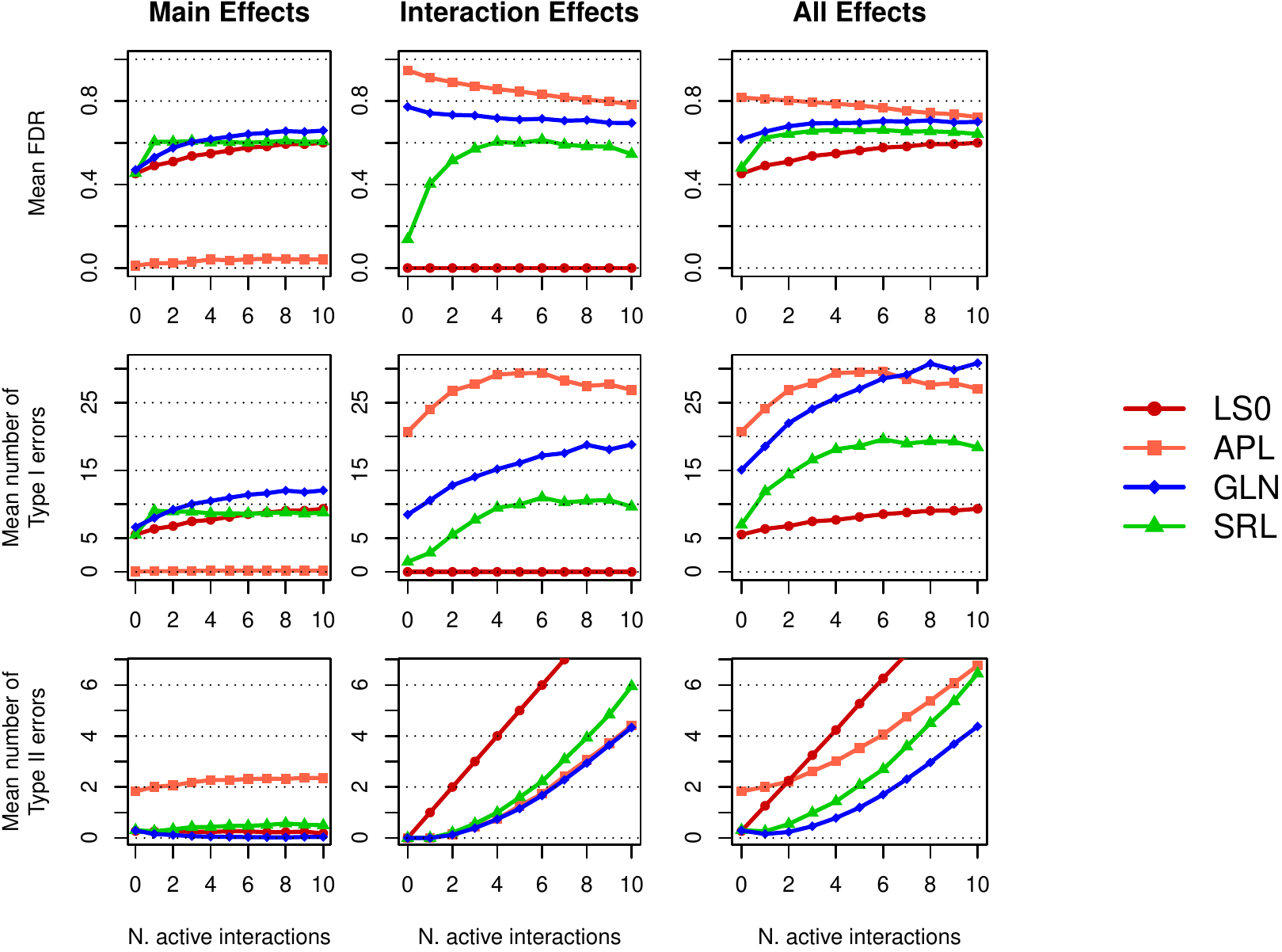} 

}

\caption{Model selection performance of various interaction fitting methods. The top three plots show the mean FDR across simulations, the middle three plots show the mean number of Type I errors, and the bottom three plots show the mean number of Type II errors. The metrics are stratified into main-effects (left), interaction effects (center), and their combined/overall values (right). LS0 refers to the lasso fit using only the original terms, APL refers to the lasso fit using the original terms and all pairwise interactions, SRL refers to the sparsity-ranked lasso fit with $\gamma = 0.5$, and GLN refers to the glinternet model. For all models, $\lambda$ was tuned with 10-fold cross-validation.}\label{fig:Fig_05}
\end{figure}

The plots in Figure \ref{fig:Fig_05} show model selection information
for each framework. When the true model has no active interactions, the
LS0 and the SRL methods look very similar in terms of FDR and the mean
number of Type I/II errors. In this same setting, the GLN and APL models
have a much higher FDR and mean number of Type I errors; this is driven
by the tendency of these models to select too many false interactions
(rendering selected models unnecessarily opaque). For all of the
generative settings, the overall FDR for the APL is very high, and it is
driven disproportionately by a high FDR in the interaction effects. The
GLN method also exhibits this differential expression of the FDR among
main and interaction effects, though to a lesser degree. This difference
is further seen in the number of Type I errors; the higher FDR in the
interaction effects translates to many more Type I errors for GLN and
APL than for SRL. The SRL method maintains approximately the same number
of Type I errors in the interaction effects and the main effects for
\(b\geq4\). This improvement in FDR/Type I error rate exhibited by SRL
compared to GLN is balanced out by a slightly higher mean number of Type
II errors, a disparity which grows with the number of active
interactions. In summary, when interactions are especially sparse, the
SRL outperforms every other method in terms of prediction and selection.
Otherwise, the SRL and GLN perform comparably to one another in terms of
prediction, although the SRL produces more interpretable/transparent
models by admitting fewer unnecessary (false) interactions.

\hypertarget{application-gene-environment-interactions}{%
\section{Application: Gene-Environment
Interactions}\label{application-gene-environment-interactions}}

\hypertarget{background}{%
\subsection{Background}\label{background}}

\begin{figure}

{\centering \includegraphics[height=3in]{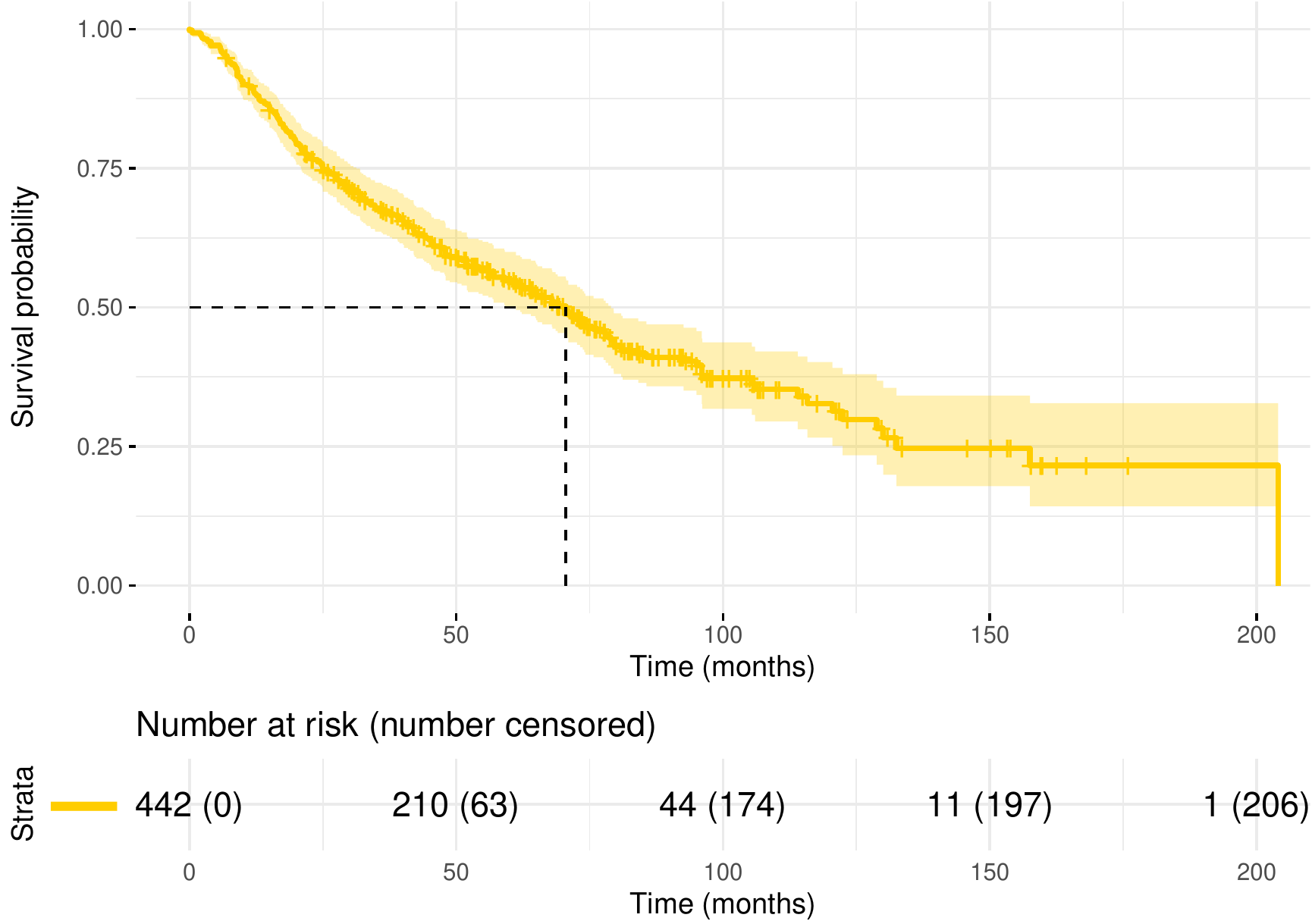} 

}

\caption{Kaplan-Meier curve for Shedden data; survival time is the primary outcome of the study for which we build regularized Cox regression models to predict.}\label{fig:Fig_06}
\end{figure}

We wish to show how SRL methods can be used in the context of genetic
data, specifically for the purpose of detecting important
gene-environment interactions. Gene-environment interactions make sense
biologically, but unfortunately, they are very difficult to detect in
practice. With high-dimensional data, the detection of any meaningful
association is sufficiently challenging, yet looking for interactions
with high-dimensional data is akin to searching for several needles in
tens of thousands of haystacks. We utilize a study that collected data
on 442 patients with lung cancer (adenocarcinoma) (Shedden et al.,
2008). For each patient, the investigators observed the time of death or
censor (the primary outcome), 22,283 gene expression measurements taken
from a sample of the lung cancer tumor, and some clinical covariates:
sex, race, age, whether the patient received chemotherapy, smoking
history, and cancer stage. The main outcome of overall survival is
presented in Figure \ref{fig:Fig_06}.

\hypertarget{methods}{%
\subsection{Methods}\label{methods}}

In order to fit models with gene-environment interactions, we take our
outcome \((\boldsymbol y, \boldsymbol d)\) as the time until
death/censor and the death indicator, respectively. We model the hazard
of death using a Cox proportional hazards model with our predictors
being comprised of the genetic and clinical covariates described in the
previous section denoted by \(X\) and \(Z\), respectively. For our
purposes, we are only interested in interactions that may occur between
\(X\) and \(Z\) or within \(Z\); we do not look for interactions that
occur within
\(X\)\footnote{Strictly speaking, the clinical covariates are not all environmental, but we decided to include all of their interactions as candidates out of interest. However, selected interactions should be considered in their proper context and accordingly labelled using appropriate terminology.}.
Since the outcome is time-to-event, we substitute the partial likelihood
for the Cox regression model into the objective function, taking the
place of the least squares term.

We investigate the performance of three candidate modeling frameworks:
the lasso using only main effects (LS), the sparsity-ranked lasso (SRL),
and the all-pairwise lasso (APL). Within the SRL framework, we set
\(\gamma = 0.5\), and investigate three different penalty schema. Since
our features consist of both clinical and genetic covariates, we treat
these as two separate covariate groups of different sizes in a model we
abbreviate as SR0, which does not include any interactions (but still
uses the SRL framework). SR1 refers to the SRL approach for both the
main and interaction effects with proportional weighting. Finally, SR2
refers to the cumulative SRL, wherein the penalty increases cumulatively
for clinical covariates, genetic covariates, and their interactions (in
that order).

The first step in the modeling process is to split the data
\(\mathbb X = \left[\boldsymbol y, \boldsymbol d, X, Z \right]\)
randomly into a training set \(\mathbb X_{\text{train}} \ (n = 342)\)
and a test set \(\mathbb X_{\text{test}} \ (n = 100)\). Second, based on
\(\mathbb X_{\text{train}}\), we use repeated (\(r=10\))
cross-validation (\(k=10\)) to tune each of the aforementioned models
(with respect to \(\lambda\)). At this stage, we also select the optimal
modeling structure. Third, we use the optimally tuned model within each
modeling structure to predict outcomes on the test set
\(\mathbb X_{\text{test}}\), comparing performance between the models
and confirming that the optimal structure we selected in the prior step
performed the best on \(\mathbb X_{\text{test}}\). Finally, we re-fit
and re-tune the optimal modeling structure using the full data
\(\mathbb X\) in order to interpret the best final model.

In order to assess predictive efficacy for cross validation, we use the
expected extra-sample Cox partial deviance, estimated as described in
the \texttt{ncvreg} documentation (Breheny and Huang, 2011). While this
measure is difficult to interpret in an absolute sense, it can be
effective in assessing predictive accuracy in a relative sense. We also
calculate an estimate of the out-of-sample \(R^2\) based on the
deviance, and we measure both accuracy measures using the test set as
well as the CV process.

As a final purely visual assessment of predictive performance, we
categorize individuals from the test data into three categories based on
their expected risk score (low-risk, medium-risk, or high-risk). The cut
points are set to be the 33rd and 67th percentile of the linear
predictions on the test set, which could vary across methods. Then,
using the test set, we plot Kaplan-Meier (KM) curves for each method
stratified by the test set's predicted risk score categories. More
separation among those stratifications on the KM plot means better
predictive performance; such delineation indicates that the model is
doing a good job of classifying high-, medium-, and low-risk patients in
the test set.

\hypertarget{results}{%
\subsection{Results}\label{results}}

\begin{figure}

{\centering \includegraphics[height=4in]{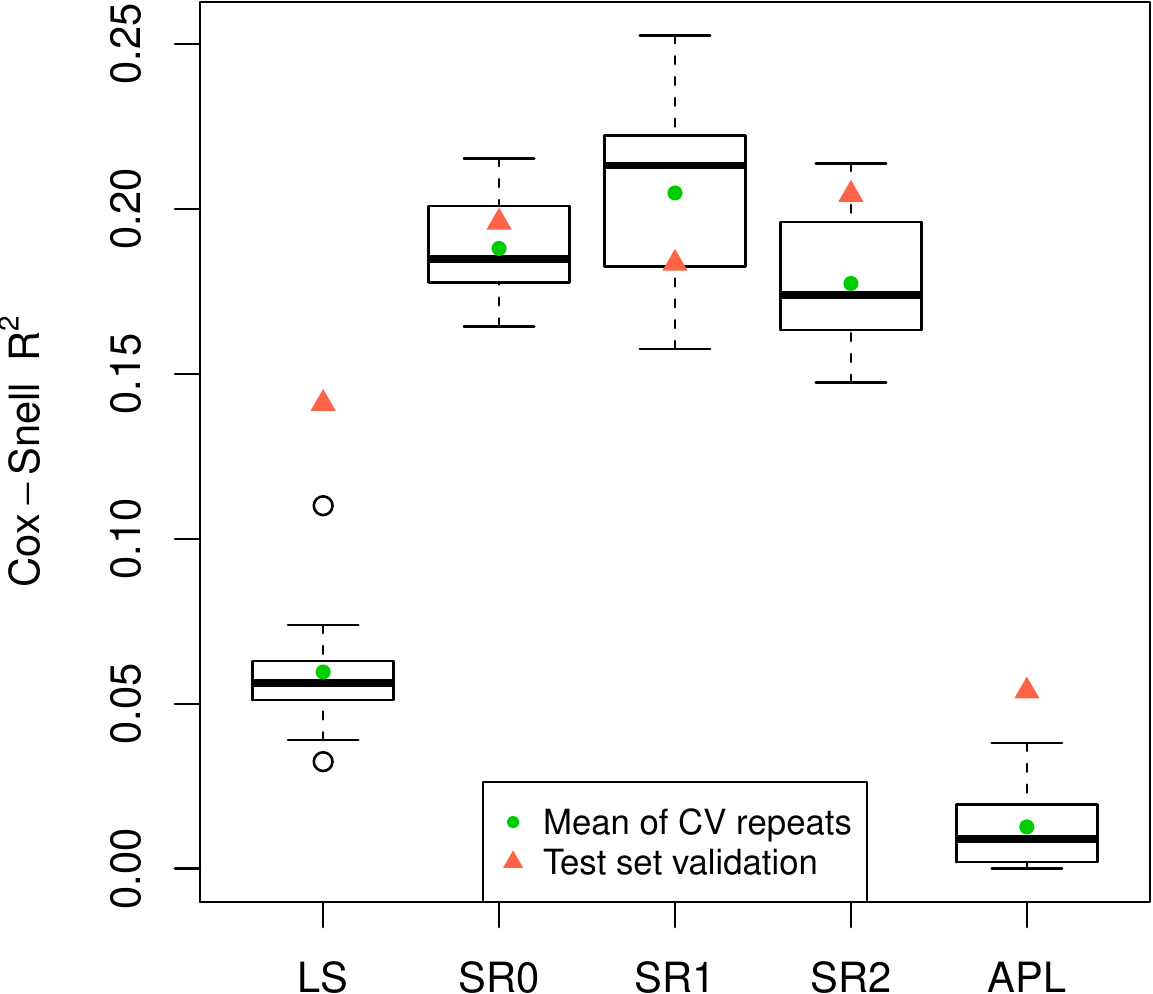} 

}

\caption{Maximum Cox-Snell $R^2$ achieved for each model. Box plots consist of the fold-averaged estimate for each of 10 repeats, so the spread of these results are due differences in the fold assignments related to the random number generation seed, and do not represent the variability in the CV estimate itself.}\label{fig:Fig_07}
\end{figure}

The estimated extra- and out-of-sample Cox partial deviance and
Cox-Snell \(R^2\) by model is shown in Figure \ref{fig:Fig_07} and Table
\ref{tab:tab02_1}. We find APL performed quite poorly, which indicates
that the consideration of pairwise interactions, without accounting for
ranked sparsity, is not a good idea. The LS method performed only
slightly better, which indicates that penalizing the genetic and
clinical covariates equally may not be advised either. The relatively
strong performance of SR0, SR1, and SR2 indicates that the
sparsity-ranked lasso achieves a satisfactory middle ground. Since these
SRL models all perform similarly, and none of them select any
interactions (see Table \ref{tab:tab02_2}), we have little to no
evidence that any prominent gene-environment interactions are capable of
being discovered.

\captionsetup{width=.82\textwidth}

\begin{table}

\caption{\label{tab:tab02_1}Estimated predictive performance (mean deviance and Cox-Snell $R^2$) calculated using extra- and out-of-sample data broken down by modeling framework. Cross-validated values are estimated with 10-folds and 10 repeats.}
\centering
\begin{tabular}[t]{lrrrr}
\toprule
\multicolumn{1}{c}{ } & \multicolumn{2}{c}{Cross-Validation} & \multicolumn{2}{c}{Test set} \\
\cmidrule(l{3pt}r{3pt}){2-3} \cmidrule(l{3pt}r{3pt}){4-5}
  & Deviance & $R^2$ & Deviance & $R^2$\\
\midrule
Lasso with only main effects (LS) & 10.35 & 0.060 & 7.77 & 0.141\\
SRL with only main effects, proportional weights (SR0) & 10.20 & 0.188 & 7.70 & 0.196\\
SRL with interactions, proportional weights (SR1) & 10.18 & 0.205 & 7.72 & 0.184\\
SRL with interactions, cumulative weights (SR2) & 10.22 & 0.177 & 7.69 & 0.204\\
All-pairwise lasso (APL) & 10.40 & 0.013 & 7.86 & 0.054\\
\bottomrule
\end{tabular}
\end{table}

\captionsetup{width=.8\textwidth}

\begin{figure}

{\centering \includegraphics[height=4.5in]{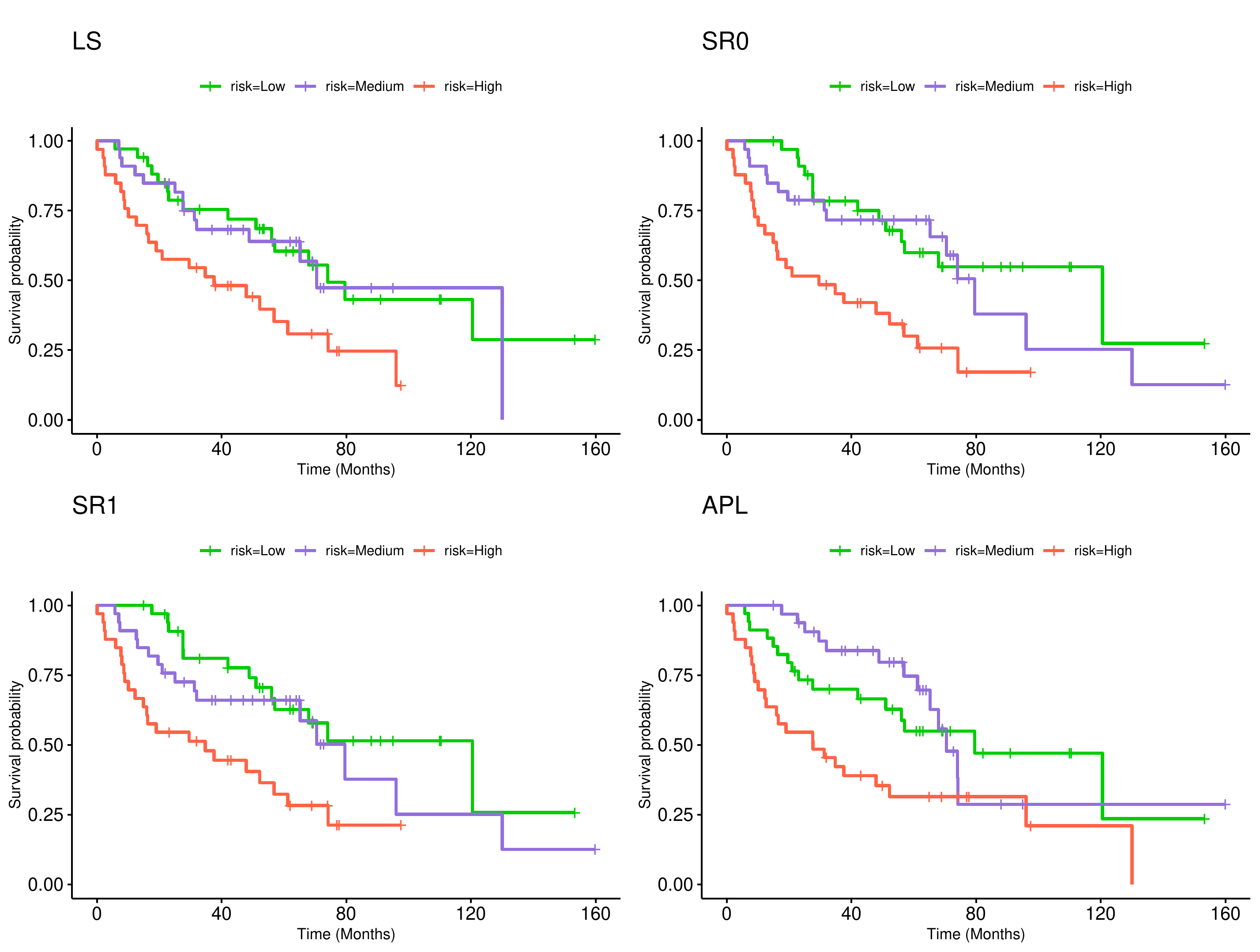} 

}

\caption{Risk score classification performance efficacy of each modeling framework using the test data set. More separation among stratifications on the KM plot signifies indicates that the model is doing a good job of classifying high-, medium-, and low-risk patients. LS refers to the lasso on the original covariates, SR0 refers to the SRL with proportional penalties on clinical and genetic covariates, SR1 refers to the SRL with interactions and proportional penalty weights, and APL refers to the all-pairwise lasso.}\label{fig:Fig_8}
\end{figure}

In Figure \ref{fig:Fig_8}, we show the categorization efficacy of each
model using the test data set. SR2 is omitted here because its
performance is very similar to SR0. In the plots, we note that the LS
and the APL models did a good job classifying high-risk patients, but
did not distinguish well between medium- and low-risk patients. The SR0
and SR1 models seem to have done a relatively good job classifying
individuals in the test set, which is most likely due to the handling of
the clinical covariates (SRL is shrinking the clinical variables
relatively less than the LS model).

After re-fitting and re-tuning each model to the entire data set, we
examine the number of selections (S) and the sum of the magnitude of the
standardized coefficients by covariate group for each optimally tuned
model in Table \ref{tab:tab02_2}. Evidently, the LS model found most of
its signal from the genetic covariates; 43 of which had nonzero
coefficients. Only 3 clinical covariates were selected, and the combined
magnitude of the standardized coefficients (\(||\beta||_1\)) was only
\(0.134\). SR0, on the other hand, found the majority of the signal to
lie in the six clinical covariates (\(||\beta||_1 = 0.938\)), though it
still found a good amount of signal (\(0.857\)) in 42 of the genetic
covariates. Neither SR1 nor SR2 selected any gene-environment
interactions. SR1 found less signal in the genetic variables than SR0 --
this indicates that the addition of interaction terms necessitated a
higher amount of shrinkage in the main effects (particularly the genetic
main effects). For SR2 however, since the coefficients are being
penalized in a cumulative fashion, the amount of signal is very similar
to SR0 when no interactions were considered. APL, although having
discovered 5 gene-environment interactions, is clearly not able to find
much signal at all; it is shrinking all of the effects considerably.
These results taken together indicate that there are no informative
gene-environment interactions, and that the clinical variables should be
penalized proportionally less than the genetic variables.

\captionsetup{width=.78\textwidth}

\begin{table}

\caption{\label{tab:tab02_2}The number of selections (S) and the sum of the magnitude of the standardized coefficients by covariate group for each (optimally tuned) model. Tuning of $\lambda$ was accomplished with 10-fold cross-validation with 10 repeats, and $\gamma$ was set to 0.5. LS refers to the lasso on the original covariates, SR0 refers to the SRL with proportional penalties on clinical and genetic covariates, SR1 refers to the SRL with interactions and proportional penalty weights, SR2 refers to the SRL with interactions and cumulative penalty weights, and APL refers to the all-pairwise lasso.}
\centering
\fontsize{10.5}{12.5}\selectfont
\begin{tabular}[t]{>{\raggedright\arraybackslash}p{2.4cm}rrrrrrrrrr}
\toprule
\multicolumn{1}{c}{  } & \multicolumn{2}{c}{LS} & \multicolumn{2}{c}{SR0} & \multicolumn{2}{c}{SR1} & \multicolumn{2}{c}{SR2} & \multicolumn{2}{c}{APL} \\
\cmidrule(l{3pt}r{3pt}){2-3} \cmidrule(l{3pt}r{3pt}){4-5} \cmidrule(l{3pt}r{3pt}){6-7} \cmidrule(l{3pt}r{3pt}){8-9} \cmidrule(l{3pt}r{3pt}){10-11}
  & S & $||\beta||_1$ & S & $||\beta||_1$ & S & $||\beta||_1$ & S & $||\beta||_1$ & S & $||\beta||_1$\\
\midrule
Clinical & 3 & 0.134 & 6 & 0.938 & 6 & 0.926 & 6 & 0.935 & 1 & 0.013\\
Genetic & 43 & 1.133 & 42 & 0.857 & 31 & 0.583 & 42 & 0.786 & 6 & 0.229\\
Env-Env & 0 & 0.000 & 0 & 0.000 & 0 & 0.000 & 0 & 0.000 & 0 & 0.000\\
Gene-Env & 0 & 0.000 & 0 & 0.000 & 0 & 0.000 & 0 & 0.000 & 5 & 0.041\\
\bottomrule
\end{tabular}
\end{table}

\captionsetup{width=.8\textwidth}

In Table \ref{tab:tab02_3}, we show how many selected variables were
shared for each model selection method. There was very high agreement in
the SR models, and in fact perfect agreement between SR0 and SR2 (they
selected all of the same variables). SR1 selected only one covariate
that was not selected by SR0 or SR2, a feature called ``checkpoint
kinase 1,'' although its estimated coefficient was very small.

\captionsetup{width=.382\textwidth}

\begin{table}

\caption{\label{tab:tab02_3}Number of selected coefficients common among each method.}
\centering
\begin{tabular}[t]{lccccc}
\toprule
  & LS & SR0 & SR1 & SR2 & APL\\
\midrule
LS & 46 & 34 & 27 & 34 & 7\\
SR0 &  & 48 & 36 & 48 & 7\\
SR1 &  &  & 37 & 36 & 7\\
SR2 &  &  &  & 48 & 7\\
APL &  &  &  &  & 12\\
\bottomrule
\end{tabular}
\end{table}

\captionsetup{width=.8\textwidth}

Finally, we will interpret the SR0 model, which was very similar to the
SR2 model. In terms of the estimated hazard ratios (HRs), the most
protective effect we found was for those in the ``never smoked'' group
(HR = 0.74). We found two clinically significant protective gene
expressions: FAM117A (HR = 0.89), and CTAGE5 (HR = 0.90). We found
harmful effects if subjects were white (HR = 1.33), male (HR = 1.32), or
had chemotherapy (HR = 1.94). Additionally for every 10 year increase in
age, the hazard increases by a multiplicative factor of 1.44.
Interestingly, the clinical coefficients in this model are similar to
the estimates from the model with only clinical covariates. Oddly, the
identified important gene expressions selected by SR0 were not used as
classifiers in the original paper. Note that since this was not a
randomized controlled trial, these effects are not indicative of causal
relationships; in particular, the high HR on chemotherapy status does
not indicate that chemotherapy was harmful.

\hypertarget{discussion}{%
\section{Discussion}\label{discussion}}

\hypertarget{strengths-and-weaknesses-of-the-srl}{%
\subsection{Strengths and Weaknesses of the
SRL}\label{strengths-and-weaknesses-of-the-srl}}

We have shown that the sparsity-ranked lasso performs relatively well
for selecting transparent models. Whereas other methods for selecting
polynomials and/or interactions tend to select overly opaque models
(models with high-order relationships that are difficult to interpret),
SRL naturally selects models that have more main effects and fewer
``complicating'' terms. In other words, the SRL limits the tendency to
select too many interactions, and controls the number of false
discoveries among interactions to be close to the same or less than that
in the main effects. Therefore, the SRL is a technique that can be
utilized and trusted to select from interactions and polynomials without
yielding overly convoluted interpretations.

Since many authors have already contributed to the problem of selecting
from all possible interactions, discussion of the SRL compared to these
competing methods is warranted. One major benefit to the SRL is that it
can be applied to survival outcomes; at the time of writing, all of the
competing methods we have mentioned are supported by open-source
software packages, but none can handle survival outcomes to our
knowledge. Thanks to the versatility of the \texttt{ncvreg} package, the
SRL method can be used for binomial, continuous, survival, or Poisson
outcomes (Breheny and Huang, 2011). Further, in \texttt{sparseR},
sparsity-ranked versions of non-convex regularization methods such as
the Minimax Concave Penalty (MCP) (Zhang, 2010) and the Smoothly Clipped
Absolute Deviations (SCAD) penalty (Fan and Li, 2001) are also feasible
and implementable. We have shown that these non-convex methods work
quite well (empirically) in this paper's supplement (Figures S7, S8).
Another benefit of the SRL to consider is the computational speed;
glinternet has been shown to be 10-10000 times faster than hierNet, and
yet our method is quite a bit faster than glinternet (at least for our
simulation settings, see supplemental Figures S9, S10). This speed-up
does not seem to change as the sample size increases, and it is
especially noticeable when cross-validation is employed to tune the
models.

Perhaps most importantly, we have found that SRL works better than
glinternet (in terms of prediction accuracy and the false discovery
rate) when there are no interactions or when interactions are especially
sparse. This strength suggests another important benefit to the SRL
producedure; it can be worthwhile, convenient, and straightforward to
extend the SRL to examine higher order interactions (and polynomials).
As opposed to competing methods, the SRL will not heavily inflate the
number of Type I errors in the course of such an investigation, even as
number of \(k\)-order interactions increases combinatorically with
\(k\).

One large weakness to the SRL is that it requires storage of a
potentially large matrix of interactions. However, recent advances in
the scalability of regularization algorithms such as the
\texttt{biglasso} package (Zeng and Breheny, 2021) are applicable to the
SRL as well. Another weakness that the SRL shares with other
regularization procedures is that the optimal mechanism of formal
inference is unclear. It is possible to extend recent advances in the
marginal false discovery rate (mFDR) (Breheny, 2018; Miller and Breheny,
2019) to the SRL framework, and this method is currently included in the
\texttt{sparseR} package. Yet whether or not this method is optimal for
formal inference and whether the mFDR works well for ranked-sparsity
settings remains an area of future research. Finally, one often
unaddressed issue with using regularization to search for important
interactions is that the model fit is sensitive to the choice of origin
among the covariates; in particular, the method is not invariant to
changes of location, such as centering, in the covariates. The tutorial
for \texttt{sparseR} goes more into detail about what can be done in
circumstances where the best origin location is unknown ahead of time.
One solution is to use our ranked-sparsity-based information criterion
RBIC (Peterson, 2019), which is invariant to location changes in the
covariates, to search for and select an optimal model, comparing this
fit with the estimates from the regularization procedure.

In the course of our exploration, some of our results indicate that the
SRL will not perform as well as competitors in certain situations. This
relatively poor performance was seen when using the cumulative SRL for
polynomials in settings with highly skewed covariates or when the
functional forms cannot be well-represented by polynomials, in which
circumstances other smoothers tend to work better. Also, the performance
of the SRL for interactions relative to glinternet seems to depend on
the hierarchical configuration of the generating model; glinternet can
perform slightly better than the SRL when the true model is strongly
hierarchical, whereas the SRL method performs better when the true model
is weakly or non-hierarchical. Their relative predictive performance
depends to an extent on the mix of strong, weak, and non-hierarchical
active interactions, and more research is needed to determine exactly
how and why this is the case.

Finally, while we have motivated intuitive guidelines for the selection
of \(\gamma\), future work should investigate the practicality and
utility of optimizing the choice of \(\gamma\) with respect to
predictive accuracy. We incorporated \(\gamma\) in the formulation of
the SRL for two reasons: (1) to show that the original lasso can be
written as a special case of the SRL when \(\gamma=0\), and (2) to
explore the apparent benefits to the cumulative SRL for polynomials from
additional tuning of its weighting scheme. For this reason, we performed
a minor amount of tuning for \(\gamma\) when applying the cumulative SRL
in this work (showing that BIC or cross-validation can be used). Outside
of the cumulative SRL, e.g.~in our real data analysis, we opted for
fixing \(\gamma=0.5\) for simplicity and because we intended each
covariate group to contribute the same amount of prior information for
SR0 and SR1. For SR2, we acknowledge that further tuning of \(\gamma\)
may yield slightly better results, but would then be less comparable to
SR0 and SR1.

\hypertarget{other-applications-of-the-srl}{%
\subsection{Other Applications of the
SRL}\label{other-applications-of-the-srl}}

Though not the primary focus of this paper, the SRL has wide reaching
applications outside of interaction and polynomial feature selection. We
have developed SRL methods for automated autoregressive (AR) order
selection for time series data (Miller et al., 2019), finding the
procedure particularly helpful for seasonal time series with uncertainty
in the seasonal period. Additionally, we have utilized the SRL in
conjunction with adaptive out-of-sample time series regression methods
to incorporate past states of a model into the fitting of current or
future models via varying penalization weights. Finally, we have
extended the SRL into what we call ``ranked cost'' contexts, wherein
candidate covariates have quantifiably different costs of data
collection. In this setting, if any correlation exists among these
features, the SRL can simultaneously optimize for predictive accuracy
and the costs of future data collection; it produces the least costly
model that can still predict as well as the optimal model. Our
exploration of these extensions is still ongoing, but initial results
have been promising.

In the context of gene-environment interactions, single nucleotide
polymorphism (SNP) data are frequently used rather than gene expression
data. SNP data are ordinal/categorical, and tend to have very high
amounts of correlation. While further exploration of the performance of
the SRL in the context of ordinal/categorical covariate data is
warranted, we postulate that ranked sparsity methods would be a fruitful
approach. In particular, ranked sparsity can be paired with other
existing regularization approaches that work well for highly correlated
data such as the elastic net, which is implemented in the
\texttt{sparseR} package. In these highly-correlated settings,
additional (minor) tuning of the SRL between
\(\gamma \in \{0, 0.5, \infty\}\) can improve its relative performance
to glinternet and the APL (Figures
S2-S6)\footnote{This minor tuning simply involves picking the best performing model between the APL ($\gamma=0$), SRL ($\gamma=0.5$), and LS0 ($\gamma=\infty$); no additional CV is necessary.}.
Further, for SNP data, a common simplifying assumption is that each SNP
count relates linearly to the outcome (treating these covariates as
numeric rather than categorical). The cumulative SRL approach for
polynomials can navigate a powerful middle-ground between these
qualitative/quantitative extremes. While it treats the SNP count as
numeric and \emph{prefers} linearity, when strong evidence of
nonlinearity is observable, the cumulative SRL approach will guardedly
introduce polynomial terms to the active feature set. Conversely, in the
absence of such evidence, the approach will not yield an overabundance
of false discoveries.

\hypertarget{conclusion}{%
\subsection{Conclusion}\label{conclusion}}

The ranked sparsity framework implements a broader definition of Occam's
Razor where a model's simplicity is not purely equated to parsimony; it
is also tied to the model's transparency and interpretability. The
sparsity-ranked lasso provides an effective and fast approach for
selecting from derived variables such as interactions or polynomials. As
opposed to other methods of interaction selection, the SRL does not
select an unreasonable number of false interaction effects and it does
not overly shrink the main effects.

\singlespacing

\hypertarget{references}{%
\section*{References}\label{references}}
\addcontentsline{toc}{section}{References}

\hypertarget{refs}{}
\leavevmode\hypertarget{ref-aic}{}%
Akaike, H. (1974) A new look at the statistical model identification.
\emph{IEEE Transactions on Automatic Control}, \textbf{19}, 716--723.

\leavevmode\hypertarget{ref-bien2013}{}%
Bien, J., Taylor, J. and Tibshirani, R. (2013) A lasso for hierarchical
interactions. \emph{The Annals of Statistics}, \textbf{41}, 1111--1141.

\leavevmode\hypertarget{ref-mbic}{}%
Bogdan, M., Frommlet, F., Biecek, P., Cheng, R., Ghosh, J.K. and Doerge,
R. (2008) Extending the modified Bayesian information criterion (mBIC)
to dense markers and multiple interval mapping. \emph{Biometrics},
\textbf{64}, 1162--1169.

\leavevmode\hypertarget{ref-boulesteix2017ipf}{}%
Boulesteix, A.-L., De Bin, R., Jiang, X. and Fuchs, M. (2017) IPF-lasso:
Integrative-penalized regression with penalty factors for prediction
based on multi-omics data. \emph{Computational and Mathematical Methods
in Medicine}, \textbf{2017}, 7691937.

\leavevmode\hypertarget{ref-mfdr1}{}%
Breheny, P.J. (2018) Marginal false discovery rates for penalized
regression models. \emph{Biostatistics}, \textbf{20}, 299--314.

\leavevmode\hypertarget{ref-breheny2011}{}%
Breheny, P. and Huang, J. (2011) Coordinate descent algorithms for
nonconvex penalized regression, with applications to biological feature
selection. \emph{Annals of Applied Statistics}, \textbf{5}, 232--253.

\leavevmode\hypertarget{ref-chenchen2008}{}%
Chen, J. and Chen, Z. (2008) Extended Bayesian information criteria for
model selection with large model spaces. \emph{Biometrika}, \textbf{95},
759--771.

\leavevmode\hypertarget{ref-chipman1996}{}%
Chipman, H. (1996) Bayesian variable selection with related predictors.
\emph{The Canadian Journal of Statistics / La Revue Canadienne de
Statistique}, \textbf{24}, 17--36.

\leavevmode\hypertarget{ref-choi2010}{}%
Choi, N.H., Li, W. and Zhu, J. (2010) Variable selection with the strong
heredity constraint and its oracle property. \emph{Journal of the
American Statistical Association}, \textbf{105}, 354--364.

\leavevmode\hypertarget{ref-scad}{}%
Fan, J. and Li, R. (2001) Variable selection via nonconcave penalized
likelihood and its oracle properties. \emph{Journal of the American
Statistical Association}, \textbf{96}, 1348--1360.

\leavevmode\hypertarget{ref-friedman2010note}{}%
Friedman, J., Hastie, T. and Tibshirani, R. (2010) A note on the group
lasso and a sparse group lasso. preprint: \emph{arXiv:1001.0736}.

\leavevmode\hypertarget{ref-hao2018}{}%
Hao, N., Feng, Y. and Zhang, H.H. (2018) Model selection for
high-dimensional quadratic regression via regularization. \emph{Journal
of the American Statistical Association}, \textbf{113}, 615--625.

\leavevmode\hypertarget{ref-aicc}{}%
Hurvich, C.M. and Tsai, C.-L. (1989) Regression and time series model
selection in small samples. \emph{Biometrika}, \textbf{76}, 297--307.

\leavevmode\hypertarget{ref-prioritylasso}{}%
Klau, S., Jurinovic, V., Hornung, R., Herold, T. and Boulesteix, A.-L.
(2018) Priority-lasso: A simple hierarchical approach to the prediction
of clinical outcome using multi-omics data. \emph{BMC bioinformatics},
\textbf{19}, 322.

\leavevmode\hypertarget{ref-recipes}{}%
Kuhn, M. and Wickham, H. (2019) recipes: Preprocessing tools to create
design matrices. R package, available at
https://CRAN.R-project.org/package=recipes.

\leavevmode\hypertarget{ref-lim2015}{}%
Lim, M. and Hastie, T. (2015) Learning interactions via hierarchical
group-lasso regularization. \emph{Journal of Computational and Graphical
Statistics}, \textbf{24}, 627--654.

\leavevmode\hypertarget{ref-mallow}{}%
Mallows, C.L. (1973) Some comments on Cp. \emph{Technometrics},
\textbf{15}, 661--675.

\leavevmode\hypertarget{ref-mfdr2}{}%
Miller, R.E. and Breheny, P. (2019) Marginal false discovery rate
control for likelihood-based penalized regression models.
\emph{Biometrical Journal}, \textbf{61}, 889--901.

\leavevmode\hypertarget{ref-statepi_miller}{}%
Miller, A.C., Peterson, R.A., Singh, I., Pilewski, S. and Polgreen, P.M.
(2019) Improving State-Level Influenza Surveillance by Incorporating
Real-Time Smartphone-Connected Thermometer Readings Across Different
Geographic Domains. \emph{Open Forum Infectious Diseases}, \textbf{6}.

\leavevmode\hypertarget{ref-dissertation}{}%
Peterson, R.A. (2019) \emph{Ranked Sparsity: A Regularization Framework
for Selecting Features in the Presence of Prior Informational
Asymmetry}. PhD thesis, Department of Biostatistics, University of Iowa.

\leavevmode\hypertarget{ref-bestNormalize}{}%
Peterson, R.A. (2021) Finding Optimal Normalizing Transformations via
bestNormalize. \emph{The R Journal}, \textbf{13}, 310--329.

\leavevmode\hypertarget{ref-orqpaper}{}%
Peterson, R.A. and Cavanaugh, J.E. (2020) Ordered quantile
normalization: A semiparametric transformation built for the
cross-validation era. \emph{Journal of Applied Statistics}, \textbf{47},
2312--2327.

\leavevmode\hypertarget{ref-rcore}{}%
R Core Team. (2020) \emph{R: A Language and Environment for Statistical
Computing}. R Foundation for Statistical Computing, Vienna, Austria.

\leavevmode\hypertarget{ref-bic}{}%
Schwarz, G. (1978) Estimating the dimension of a model. \emph{The Annals
of Statistics}, \textbf{6}, 461--464.

\leavevmode\hypertarget{ref-shedden2008}{}%
Shedden, K., Taylor, J., Enkemann, S., Tsao, M., Yeatman, T., Gerald,
W., et al. (2008) Gene expression-based survival prediction in lung
adenocarcinoma: A multi-site, blinded validation study. \emph{Nature
Medicine}, \textbf{14}, 822--827.

\leavevmode\hypertarget{ref-biomarkertreatmentinteractions}{}%
Ternès, N., Rotolo, F., Heinze, G. and Michiels, S. (2017)
Identification of biomarker-by-treatment interactions in randomized
clinical trials with survival outcomes and high-dimensional spaces.
\emph{Biometrical Journal}, \textbf{59}, 685--701.

\leavevmode\hypertarget{ref-tibs1996}{}%
Tibshirani, R. (1996) Regression shrinkage and selection via the lasso.
\emph{Journal of the Royal Statistical Society: Series B}, \textbf{58},
267--288.

\leavevmode\hypertarget{ref-adaptivelsglm}{}%
Wang, M. and Wang, X. (2014) Adaptive lasso estimators for ultrahigh
dimensional generalized linear models. \emph{Statistics \& Probability
Letters}, \textbf{89}, 41--50.

\leavevmode\hypertarget{ref-mgcv}{}%
Wood, S.N. (2011) Fast stable restricted maximum likelihood and marginal
likelihood estimation of semiparametric generalized linear models.
\emph{Journal of the Royal Statistical Society (B)}, \textbf{73}, 3--36.

\leavevmode\hypertarget{ref-zeng2017}{}%
Zeng, Y. and Breheny, P. (2021) The biglasso Package: A Memory- and
Computation-Efficient Solver for Lasso Model Fitting with Big Data in R.
\emph{The R Journal}, \textbf{12}, 6--19.

\leavevmode\hypertarget{ref-mcp}{}%
Zhang, C.-H. (2010) Nearly unbiased variable selection under minimax
concave penalty. \emph{The Annals of Statistics}, \textbf{38}, 894--942.

\leavevmode\hypertarget{ref-adaptivelasso}{}%
Zou, H. (2006) The adaptive lasso and its oracle properties.
\emph{Journal of the American Statistical Association}, \textbf{101},
1418--1429.

\end{document}